\documentclass[twocolumn]{aastex61}

\usepackage[T1]{fontenc}
\usepackage[utf8]{inputenc}
\usepackage{amsmath}

\usepackage[noend]{algpseudocode}
\newcommand*\Let[2]{\State #1 $\gets$ #2}
\algrenewcommand\algorithmicrequire{\textbf{Precondition:}}
\algrenewcommand\algorithmicensure{\textbf{Postcondition:}}

\usepackage{tablefootnote}

\renewcommand{\b}{\pmb}
\newcommand{\tbf}{\textbf}
\renewcommand{\d}{{\rm d}}
\newcommand{\atikh}{\b{\hat a}_\text{Tikhonov}}

\newcommand{\astrocaltech}{
    \affiliation{
        Department of Astronomy,
        California Institute of Technology,
        1200 E California Blvd,
        Pasadena, CA 91125, USA
    }
}

\newcommand{\eecaltech}{
    \affiliation{
        Department of Electrical Engineering,
        California Institute of Technology,
        1200 E California Blvd,
        Pasadena, CA 91125, USA
    }
}

\newcommand{\ovro}{
    \affiliation{
        California Institute of Technology,
        Owens Valley Radio Observatory,
        Big Pine, CA 93513, USA
    }
}

\newcommand{\jpl}{
    \affiliation{
        Jet Propulsion Laboratory,
        California Institute of Technology,
        4800 Oak Grove Dr,
        Pasadena, CA 91109, USA
    }
}

\newcommand{\nrao}{
    \affiliation{
        National Radio Astronomy Observatory,
        P.O. Box O,
        Socorro, NM 87801, USA
    }
}

\newcommand{\harvard}{
    \affiliation{
        Harvard-Smithsonian Center for Astrophysics,
        60 Garden Street,
        Cambridge, MA 02138, USA
    }
}

\newcommand{\berkeley}{
    \affiliation{
        University of California Berkeley,
        501 Campbell Hall,
        Berkeley, CA 94720, USA
    }
}

\newcommand{\swinburne}{
    \affiliation{
        Centre for Astrophysics \& Supercomputing,
        Swinburne University of Technology,
        PO Box 218,
        Hawthorn, VIC 3122, Australia
    }
}

\newcommand{\unm}{
    \affiliation{
        Department of Physics and Astronomy,
        University of New Mexico,
        Albuquerque, NM 87131, USA
    }
}

\newcommand{\vt}{
    \affiliation{
        Bradley Department of Electrical \& Computer Engineering,
        Blacksburg, VA 24061, USA
    }
}

\newcommand{\nvidia}{
    \affiliation{
        NVIDIA Corporation,
        2701 San Tomas Expressway,
        Santa Clara, CA 95050, USA
    }
}

\newcommand{\chalmers}{
    \affiliation{
        Department of Space, Earth and Environment,
        Chalmers University of Technology,
        Onsala Space Observatory,
        43992 Onsala, Sweden
    }
}

\newcommand{\google}{
    \affiliation{
        Google,
        340 Main St,
        Venice, CA 90291, USA
    }
}

\begin{document}


\title{
    The Radio Sky at Meter Wavelengths: $m$-Mode Analysis Imaging\\
    with the Owens Valley Long Wavelength Array
}

\author{Michael~W.~Eastwood} \astrocaltech
\author{Marin~M.~Anderson}   \astrocaltech
\author{Ryan~M.~Monroe}      \eecaltech
\author{Gregg~Hallinan}      \astrocaltech

\author{Benjamin~R.~Barsdell} \harvard \nvidia
\author{Stephen~A.~Bourke}    \astrocaltech \chalmers
\author{M.~A.~Clark}          \nvidia
\author{Steven~W.~Ellingson}  \vt
\author{Jayce~Dowell}         \unm
\author{Hugh~Garsden}         \harvard
\author{Lincoln~J.~Greenhill} \harvard
\author{Jacob~M.~Hartman}     \google
\author{Jonathon~Kocz}        \astrocaltech
\author{T.~Joseph~W.~Lazio}   \jpl
\author{Danny~C.~Price}       \harvard \berkeley \swinburne
\author{Frank~K.~Schinzel}    \nrao \unm
\author{Gregory~B.~Taylor}    \unm
\author{Harish~K.~Vedantham}  \astrocaltech
\author{Yuankun~Wang}         \astrocaltech
\author{David~P.~Woody}       \ovro

\correspondingauthor{Michael W. Eastwood}
\email{mweastwood@astro.caltech.edu}

\begin{abstract}
    A host of new low-frequency radio telescopes seek to measure the 21-cm transition of neutral
    hydrogen from the early universe.  These telescopes have the potential to directly probe star
    and galaxy formation at redshifts $20 \gtrsim z \gtrsim 7$, but are limited by the dynamic range
    they can achieve against foreground sources of low-frequency radio emission. Consequently, there
    is a growing demand for modern, high-fidelity maps of the sky at frequencies below 200 MHz for
    use in foreground modeling and removal.  We describe a new widefield imaging technique for
    drift-scanning interferometers, Tikhonov-regularized $m$-mode analysis imaging.  This technique
    constructs images of the entire sky in a single synthesis imaging step with exact treatment of
    widefield effects.  We describe how the CLEAN algorithm can be adapted to deconvolve maps
    generated by $m$-mode analysis imaging. We demonstrate Tikhonov-regularized $m$-mode analysis
    imaging using the Owens Valley Long Wavelength Array (OVRO-LWA) by generating 8 new maps of the
    sky north of $\delta=-30^\circ$ with 15 arcmin angular resolution, at frequencies evenly spaced
    between 36.528~MHz and 73.152~MHz, and $\sim$800 mJy/beam thermal noise. These maps are a
    10-fold improvement in angular resolution over existing full-sky maps at comparable frequencies,
    which have angular resolutions $\ge 2^\circ$. Each map is constructed exclusively from
    interferometric observations and does not represent the globally averaged sky brightness. Future
    improvements will incorporate total power radiometry, improved thermal noise, and improved
    angular resolution -- due to the planned expansion of the OVRO-LWA to 2.6~km baselines.  These
    maps serve as a first step on the path to the use of more sophisticated foreground filters in
    21-cm cosmology incorporating the measured angular and frequency structure of all foreground
    contaminants.
\end{abstract}


\section{Introduction}

At redshifts $20 \gtrsim z \gtrsim 7$, the 21-cm hyperfine structure line of neutral hydrogen is
expected to produce a 10 to 100 mK perturbation in the Cosmic Microwave Background (CMB) spectrum
\citep{2006PhR...433..181F, 2012RPPh...75h6901P}. The amplitude of this perturbation on a given
line-of-sight is a function of the neutral fraction of hydrogen, the baryon overdensity, the spin
temperature relative to the CMB temperature at the given redshift, and the line-of-sight peculiar
velocity of the gas.  The spatial power spectrum of this perturbation is thought to be dominated by
inhomogeneous heating of the IGM at $z\sim 20$ \citep{2014MNRAS.437L..36F}, and by growing ionized
bubbles during the EoR at $z\sim 7$ where a detection can constrain the ionizing efficiency of early
galaxies, the UV photon mean-free-path, and the minimum halo mass that can support star formation
\citep{2015MNRAS.449.4246G}.

Current 21-cm cosmology experiments can be broadly separated into two classes: global signal
experiments that aim to detect the spectral signature of the cosmologically redshifted 21-cm
transition after averaging over the entire sky (otherwise known as the monopole), and power spectrum
experiments that incorporate angular information to attempt to measure the 3D spatial power spectrum
of cosmological 21-cm perturbations.  Ongoing global signal experiments include EDGES
\citep{2010Natur.468..796B, 2017ApJ...847...64M}, LEDA \citep{2017arXiv170909313P}, BIGHORNS
\citep{2015PASA...32....4S}, SCI-HI \citep{2014ApJ...782L...9V}, and SARAS 2
\citep{2017arXiv170306647S}.  Ongoing power spectrum experiments include PAPER/HERA
\citep{2015ApJ...809...61A, 2016arXiv160607473D}, LOFAR \citep{2017ApJ...838...65P}, and the MWA
\citep{2016ApJ...833..102B, 2016MNRAS.460.4320E}.

Just as for CMB experiments, foreground removal or suppression is an essential component of both
classes of 21-cm cosmology experiments. The brightness temperature of the galactic synchrotron
emission at high galactic latitudes is measured by \citet{2017MNRAS.464.4995M} as
\begin{equation}
    T \sim 300\,{\rm K} \times \left(\frac{\nu}{150\,{\rm MHz}}\right)^{-2.6}\,.
\end{equation}
Therefore, experiments conservatively need to achieve 5 orders of dynamic range against this
foreground emission before the cosmological signal can be measured. Current foreground removal
methods (for example, \citealt{2012ApJ...756..165P}) rely on the assumption that the foreground
emission is spectrally smooth. The low-frequency radio sky is composed of several components:
galactic synchrotron emission, supernova remnants, radio galaxies, free-free emission and absorption
from \ion{H}{2} regions, and a confusing background of radio sources.  Ideally, a foreground removal
strategy should be informed by the measured spatial structure and frequency spectrum of all
foreground components. For instance, CMB experiments typically construct several maps at several
frequencies to enable component separation.  At low frequencies, this possibility is limited by the
availability of suitable high-fidelity sky maps on angular scales ranging from tens of degrees to
arcminutes.

Recently, a host of new low-frequency sky surveys have been conducted including MSSS
\citep{2015A&A...582A.123H}, GLEAM \citep{2015PASA...32...25W}, and TGSS
\citep{2017A&A...598A..78I}. However, the primary data product generated by these surveys is a
catalog of radio point sources. At 45 MHz, \citet{2011A&A...525A.138G} created a map of the sky that
captures the diffuse emission with 5$^\circ$ resolution.  The LWA1 Low Frequency Sky Survey
\citep{2017MNRAS.469.4537D} similarly maps the sky at a range of frequencies between 35~MHz and
80~MHz with resolution between 4.5$^\circ$ and 2$^\circ$.

The Global Sky Model (GSM) \citep{2008MNRAS.388..247D} is currently the most commonly used
foreground model. The GSM is a non-parametric interpolation of various maps between 10 MHz and 100
GHz. However, the majority of information contained in the GSM is derived at frequencies $>1.4$ GHz
where the majority of the modern, high-fidelity input maps are located. At 408~MHz the venerable
Haslam map \citep{1981A&A...100..209H, 1982A&AS...47....1H} covers the entire sky at $1^\circ$
resolution.  Below 408~MHz the GSM uses 3 input sky maps. \citet{2017MNRAS.464.3486Z} construct an
improved GSM with 5 maps below 408~MHz, and \citet{2017MNRAS.469.4537D} use the LWA1 to improve the
GSM with their own sky maps.  However, the GSM generally suffers from low angular resolution ($\sim
5^\circ$) and systematic errors associated with instrumental artifacts in the input maps.  For
instance, \citet{2017MNRAS.469.4537D} report errors of $\pm 50\%$ between the GSM and their own maps
at 74 MHz, which they attribute to the increasing contribution of free-free absorption and
modifications to the synchrotron spectral index at low frequencies.

Widefield interferometric synthesis imaging is a challenging computational problem, and it has been
particularly difficult to capture large angular scales $\gg 10^\circ$ and small angular scales $\ll
1^\circ$ in a single synthesis image. We will derive a new imaging technique -- Tikhonov regularized
$m$-mode analysis imaging -- that allows a drift-scanning interferometer to image the entire visible
sky in a single coherent synthesis imaging step with no gridding and no mosaicking.

As a demonstration of this technique we apply Tikhonov regularized $m$-mode analysis imaging to the
Owens Valley Long Wavelength Array (OVRO-LWA), and generate a series of new low-frequency maps of
the sky between 36.528~MHz and 73.152~MHz.  These maps capture the full sky visible from the Owens
Valley Radio Observatory (OVRO) with angular resolution of $\sim 15$~arcmin.  These new maps
complement the existing full-sky maps at these frequencies with greatly improved angular resolution.

We aim for these maps to inform foreground removal strategies in 21-cm cosmology, and we anticipate
additional ancillary science taking advantage of the combination of high fidelity and high
resolution of these maps, including but not limited to studies of the cosmic ray emissivity at low
frequencies, searches for giant radio galaxies, and constraining the galactic synchrotron spectrum.
The maps will be made freely available online at the Legacy Archive for Microwave Background Data
Analysis (LAMBDA)\footnote{\url{https://lambda.gsfc.nasa.gov/}}.

The structure of this paper is as follows. In \S\ref{sec:imaging}, we present Tikhonov-regularized
$m$-mode analysis imaging, a new imaging technique that allows us to image the entire visible sky in
one coherent synthesis imaging step with exact widefield corrections. In \S\ref{sec:observations} we
describe our observations with the Owens Valley Long Wavelength Array (OVRO-LWA). In
\S\ref{sec:results} we present the sky maps and compare these maps against other low-frequency sky
maps.  In \S\ref{sec:error} we discuss some of the sources of error present in the maps, and finally
in \S\ref{sec:conclusion} we present our conclusions.

\section{All-Sky Imaging}\label{sec:imaging}

The goal of all imaging algorithms is to estimate the brightness of the sky $I_\nu(\hat r)$ in the
direction $\hat r$ and frequency $\nu$.  A radio interferometer measures the visibilities
$V^{ij,pq}_{\nu}$ between pairs of antennas numbered $i$ and $j$ respectively, and between
polarizations labeled $p$ and $q$ respectively. We will neglect subtleties associated with polarized
imaging, so the Stokes-$I$ visibilities are constructed from the sum of the $pp$ and $qq$
correlations such that $V^{ij}_{\nu} = (V^{ij,pp}_{\nu}+V^{ij,qq}_{\nu})/2$.  If the antennas are
separated by the baseline $\vec b_{ij}$, and $A_\nu(\hat r)$ describes an antenna's response to the
incident Stokes-$I$ radiation (here assumed to be the same for each antenna), then
\begin{equation}\label{eq:basic-imaging}
    V^{ij}_\nu = \int_\text{sky}
                 A_\nu(\hat r) I_\nu(\hat r)
                 \exp\bigg(2\pi i \hat r\cdot\vec b_{ij}/\lambda\bigg) \,\d\Omega \, ,
\end{equation}
where the integral runs over solid angle $\Omega$.  Constructing an image from the output of a radio
interferometer consists of estimating $I_\nu(\hat r)$ given the available measurements $V^{ij}_\nu$.

For later convenience we will define the baseline transfer function $B^{ij}_\nu(\hat r)$ such that
\begin{equation}\label{eq:baseline-transfer-function}
    V^{ij}_\nu = \int_\text{sky} B^{ij}_\nu(\hat r) I_\nu(\hat r) \,\d\Omega \, .
\end{equation}
The baseline transfer function defines the response of a single baseline to the sky, and is a
function of the antenna primary beam, and baseline length and orientation.

Naively one might attempt to solve Equation~\ref{eq:basic-imaging} by discretizing, and subsequently
solving the resulting matrix equation. If the interferometer is composed of $N_\text{base}$
baselines, and measures $N_\text{freq}$ frequency channels over $N_\text{time}$ integrations, then
the entire data set consists of $N_\text{base}N_\text{freq}N_\text{time}$ complex numbers. If the
sky is discretized into $N_\text{pix}$ pixels then the relevant matrix has dimensions of
$(N_\text{base}N_\text{freq}N_\text{time})\times(N_\text{pix})$. For making single-channel maps with
the OVRO-LWA this becomes a 5 petabyte array (assuming each matrix element is a 64-bit complex
floating point number).  This matrix equation is therefore prohibitively large, and solving
Equation~\ref{eq:basic-imaging} by means of discretization is usually intractable, although
\citet{2017MNRAS.465.2901Z} demonstrate this technique with the MITEOR telescope.

Instead, it is common to make mild assumptions that simplify Equation~\ref{eq:basic-imaging} and
ease the computational burden in solving for $I_\nu(\hat r)$. For example, when all of the baselines
$\vec b_{ij}$ lie in a plane and the field-of-view is small, Equation~\ref{eq:basic-imaging} can be
well-approximated by a two-dimensional Fourier transform \citep{2001isra.book.....T}. The
restriction on baseline co-planarity and field-of-view can be relaxed by using W-projection
\citep{2008ISTSP...2..647C}. Known primary beam effects can also be accounted for during imaging by
using A-projection \citep{2013ApJ...770...91B}.

\subsection{$m$-Mode Analysis}\label{sec:mmode-analysis}

Transit telescopes can take advantage of a symmetry in Equation~\ref{eq:basic-imaging} that greatly
reduces the amount of computer time required to image the full-sky with exact incorporation of
widefield imaging effects. This technique, called $m$-mode analysis, also obviates the need for
gridding and mosaicking. Instead the entire sky is imaged in one coherent synthesis imaging step.  We
will briefly summarize $m$-mode analysis below, but the interested reader should consult
\citet{2014ApJ...781...57S, 2015PhRvD..91h3514S} for a complete derivation.

In the context of $m$-mode analysis a transit telescope is any interferometer for which the response
pattern of the individual elements does not change with respect to time. This may be an
interferometer like the OVRO-LWA where the correlation elements are fixed dipoles, but it may also
be an interferometer like LOFAR or the MWA if the steerable beams are held in a fixed position (not
necessarily at zenith). The interferometer also does not necessarily have to be homogeneous.
Heterogeneous arrays composed of several different types of antennas are allowed as long as care is
taken to generalize Equation~\ref{eq:basic-imaging} for a heterogeneous array.

For a transit telescope, the visibilities $V^{ij}_\nu$ are a periodic function of sidereal
time.\footnote{
    This is not strictly true. Ionospheric fluctuations and non-sidereal sources (such as the Sun)
    will violate this assumption. This paper will, however, demonstrate that the impact on the final
    maps is mild.
}
Therefore it is a natural operation to compute the Fourier transform of the visibilities with
respect to sidereal time $\phi\in[0,2\pi)$.
\begin{equation}
    V^{ij}_{m,\nu} = \int_0^{2\pi} V^{ij}_\nu(\phi)\exp\bigg(-im\phi\bigg)\,\d\phi
\end{equation}
The output of this Fourier transform is the set of $m$-modes $V^{ij}_{m,\nu}$ where
$m=0,\,\pm1,\,\pm2,\,\ldots$ is the Fourier conjugate variable to the sidereal time. The $m$-mode
corresponding to $m=0$ is a simple average of the visibilities over sidereal time. Similarly $m=1$
corresponds to the component of the visibilities that varies over half-day timescales. Larger values
of $m$ correspond to components that vary on quicker timescales.

\citet{2014ApJ...781...57S, 2015PhRvD..91h3514S} show that there is a discrete linear relationship
between the measured $m$-modes $V^{ij}_{m,\nu}$ and the spherical harmonic coefficients of the sky
brightness $a_{lm,\nu}$.
\begin{equation}\label{eq:m-mode-sum-equation}
    V^{ij}_{m,\nu} = \sum_l B^{ij}_{lm,\nu} a_{lm,\nu}\,,
\end{equation}
where the transfer coefficients $B^{ij}_{lm,\nu}$ are computed from the spherical harmonic transform
of the baseline transfer function defined by Equation~\ref{eq:baseline-transfer-function}. These
transfer coefficients define the interferometer's response to the corresponding spherical harmonic
coefficients.

Equation~\ref{eq:m-mode-sum-equation} can be recognized as a matrix equation where the transfer
matrix $\b B$ is block-diagonal.
\begin{equation}\label{eq:m-mode-matrix-equation}
    \overbrace{\left(
        \begin{array}{c}
            \vdots \\
            m\text{-modes} \\
            \vdots \\
        \end{array}
    \right)}^{\b v}
    =
    \overbrace{\left(
        \begin{array}{ccc}
            \ddots & & \\
            & \text{transfer matrix} & \\
            & & \ddots \\
        \end{array}
    \right)}^{\b B}
    \overbrace{\left(
        \begin{array}{c}
            \vdots \\
            a_{lm} \\
            \vdots \\
        \end{array}
    \right)}^{\b a}
\end{equation}
\begin{equation}
    \b B = \left(\begin{array}{cccc}
        m = 0 &&& \\
              & m=\pm1 && \\
              && m=\pm2 & \\
              &&& \ddots \\
    \end{array}\right)
\end{equation}
The vector $\b v$ contains the list of $m$-modes and the vector $\b a$ contains the list of
spherical harmonic coefficients representing the sky brightness. In order to take advantage of the
block-diagonal structure in $\b B$, $\b v$ and $\b a$ must be sorted by the absolute value of $m$.
Positive and negative values of $m$ are grouped together because the brightness of the sky is
real-valued, and the spherical harmonic transform of a real-valued function has $a_{l(-m)} = (-1)^m
a_{lm}^*$.

In practice, we now need to pick the set of spherical harmonics we will use to represent the sky.
For an interferometer like the OVRO-LWA with many short baselines, a sensible choice is to use all
spherical harmonics with $l\le l_\text{max}$ for some $l_\text{max}$. The parameter $l_\text{max}$
is determined by the maximum baseline length of the interferometer.  For an interferometer without
short spacings, a minimum value for $l$ might also be used. This $l_\text{min}$ parameter should be
determined by the minimum baseline length. A rough estimate of $l$ for a baseline of length $b$ at
frequency $\nu$ is $l \sim \pi b\nu/c$. Based on this estimate for the OVRO-LWA and other
computational considerations, we therefore adapt $l_\text{min}=1$ and $l_\text{max}=1000$ across all
frequencies. However, this choice of $l_\text{max}$ actually limits the angular resolution above
55~MHz, and therefore future work will increase $l_\text{max}$ to obtain better angular resolution.

The interferometer's sensitivity to the monopole ($a_{00}$) deserves special consideration.
\citet{2016ApJ...826..116V} prove -- under fairly general assumptions -- that a baseline with
nonzero sensitivity to $a_{00}$ must also have some amount of cross-talk or common-mode noise.  In
fact the sensitivity to $a_{00}$ is proportional to a sum of these effects. For example, one way a
baseline can have nonzero sensitivity to $a_{00}$ is if the baseline is extremely short. In this
case the antennas are so close together that voltage fluctuations in one antenna can couple into the
other antenna. In order to make an interferometric measurement of $a_{00}$, this coupling must be
measured and calibrated. Consequently, we set $a_{00}=0$ in our analysis. In the future this
limitation will be addressed with the inclusion of calibrated total power radiometry.

The size of a typical block in the transfer matrix is
$(2N_\text{base}N_\text{freq})\times(l_\text{max})$. If each element of the matrix is stored as a
64-bit complex floating point number, a single block is 500 MB for the case of single-channel
imaging with the OVRO-LWA, which a modern computer can easily store and manipulate in memory.
However, with additional bandwidth these blocks quickly become unwieldy; thus as a first pass, the
analysis in this paper is restricted to single-channel imaging. Note also that for the OVRO-LWA
$N_\text{base} \gg l_\text{max}$, so there are more measurements than unknowns in
Equation~\ref{eq:m-mode-matrix-equation}.

The key advantage of $m$-mode analysis is the block-diagonal structure of
Equation~\ref{eq:m-mode-matrix-equation}. The computational complexity of many common matrix
operations (eg. solving a linear system of equations) is $\mathcal{O}(N^3)$.  By splitting the
equation into $N$ independent blocks, the number of floating point operations required to operate on
the full matrix is reduced by a factor of $N^2$. This computational savings is what makes this
matrix algebra approach to interferometric imaging feasible.

\subsection{$m$-Mode Analysis Imaging}\label{sec:mmode-imaging}

Imaging in $m$-mode analysis essentially amounts to inverting
Equation~\ref{eq:m-mode-matrix-equation} to solve for the spherical harmonic coefficients $\b a$.
The linear-least squares solution, which minimizes $\|\b v - \b B\b a\|^2$, is given by
\begin{equation}
    \b{\hat a}_\text{LLS} = (\b B^*\b B)^{-1}\b B^*\b v\,,
\end{equation}
where $^*$ indicates the conjugate-transpose.

However, usually one will find that $\b B$ is not full-rank, and hence $\b B^*\b B$ is not an
invertible matrix. For example, an interferometer located in the northern hemisphere will never see
a region of the southern sky centered on the southern celestial pole. The $m$-modes contained in the
vector $\b v$ must contain no information about the sky around the southern celestial pole, and
therefore the act of multiplying by $\b B$ must destroy some information about the sky. The
consequence of this fact is that $\b B$ must have at least one singular value that is equal to zero.
It then follows that $\b B^*\b B$ must have at least one eigenvalue that is equal to zero, which
means it is not an invertible matrix.

Another way of looking at the problem is that because the interferometer is not sensitive to part of
the southern hemisphere, there are infinitely many possible solutions to
Equation~\ref{eq:m-mode-matrix-equation} that will fit the measured data equally well.  We will
therefore regularize the problem and apply an additional constraint that prefers a unique yet
physically reasonable solution.

\subsubsection{Tikhonov Regularization}

The process of Tikhonov regularization minimizes $\|\b v - \b B\b a\|^2 + \varepsilon\|\b a\|^2$ for
some arbitrary value of $\varepsilon > 0$ chosen by the observer. The solution that minimizes this
expression is given by
\begin{equation}\label{eq:tikhonov-solution}
    \atikh = (\b B^*\b B + \varepsilon\b I)^{-1}\b B^*\b v\,.
\end{equation}
Tikhonov regularization adds a small value $\varepsilon$ to the diagonal of $\b B^*\b B$, fixing the
matrix's singularity. By using the singular value decomposition (SVD) of the matrix $\b B = \b U \b
\Sigma \b V^*$, Equation~\ref{eq:tikhonov-solution} becomes
\begin{equation}
    \atikh = \b V (\b\Sigma^2 + \varepsilon \b I)^{-1}\b\Sigma \b U^*\b v\,,
\end{equation}
where
\[
    \b\Sigma = \left(
        \begin{array}{ccc}
            \sigma_1 & & \\
                     & \sigma_2 & \\
                     & & \ddots \\
        \end{array}
    \right)\,.
\]
The diagonal elements of $\b\Sigma$ are the singular values of $\b B$. The contribution of each
singular component to the Tikhonov-regularized solution is scaled by $\sigma_i / (\sigma_i^2 +
\varepsilon)$, where $\sigma_i$ is the singular value for the $i$th singular component. Tikhonov
regularization therefore acts to suppress any component for which
$\sigma_i\lesssim\sqrt{\varepsilon}$.  If $\sigma_i = 0$, the component is set to zero.

In practice the measurement $\b v$ is corrupted by noise with covariance $\b N$. For illustrative
purposes we will assume that $\b N=n\b I$ for some $n>0$. In this case, the covariance of the
Tikhonov-regularized spherical harmonic coefficients is
\begin{equation}
    \b C = n \b V (\b\Sigma^2 + \varepsilon\b I)^{-2} \b\Sigma^2 \b V^*\,.
\end{equation}
Each singular component is scaled by a factor of $\sigma_i^2/(\sigma_i^2 + \varepsilon)^2$.  In the
absence of Tikhonov regularization ($\varepsilon=0$), singular components with the smallest singular
values -- the ones that the interferometer is the least sensitive to -- actually come to dominate
the covariance of the measured spherical harmonic coefficients. Tikhonov regularization improves
this situation by downweighting these components.

\subsubsection{L-Curves}

\begin{figure}[t]
    \includegraphics[width=\columnwidth]{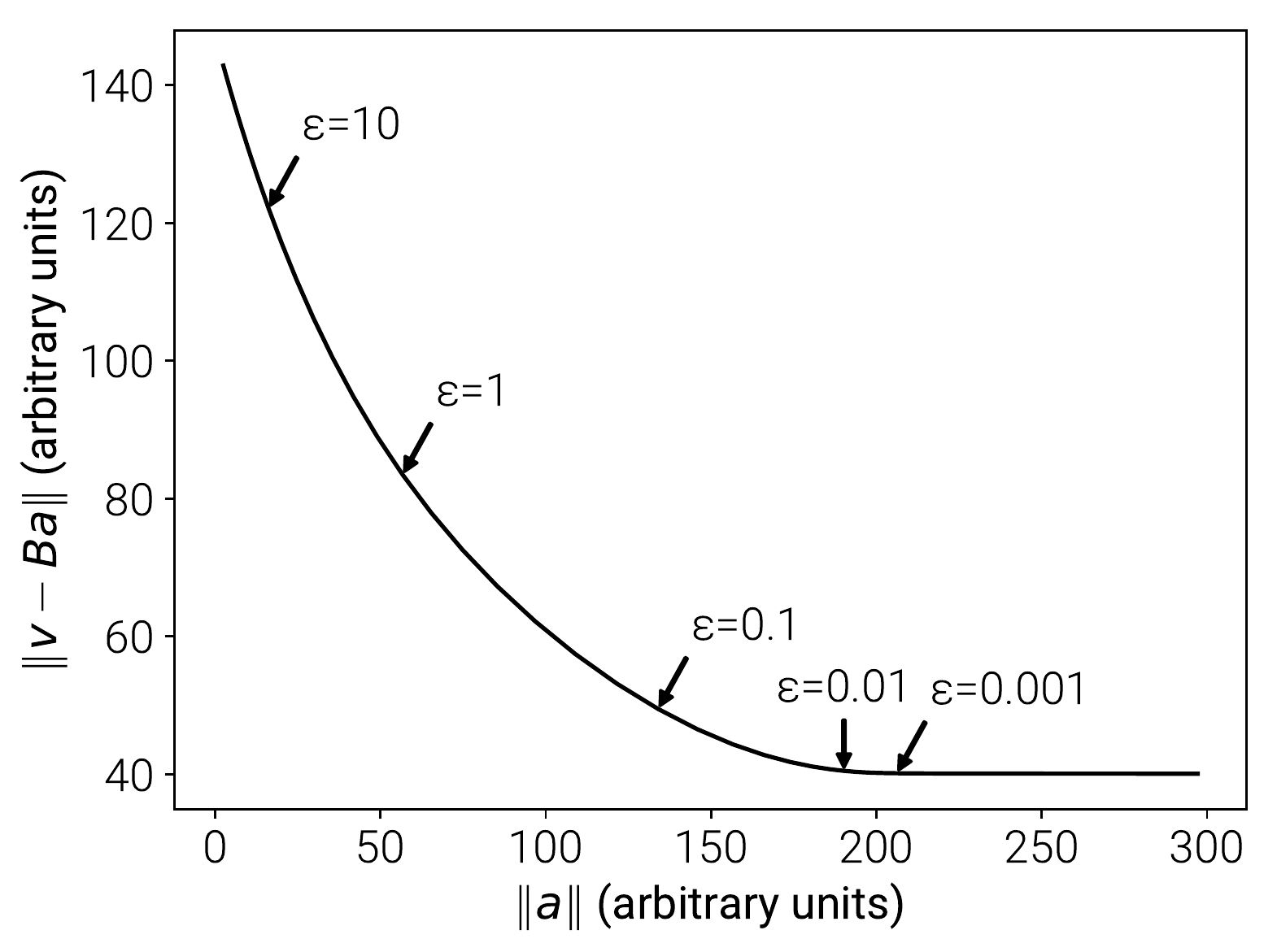}
    \caption{
        An example L-curve computed from OVRO-LWA data at 36.528~MHz by trialing 200 different
        values of the regularization parameter $\varepsilon$. The $x$-axis is the norm of the
        solution (in this case the spherical harmonic coefficients) given in arbitrary units, and
        the $y$-axis is the least-squares norm given in arbitrary units. Where the regularization
        parameter is small, the norm of the solution grows rapidly. Where the regularization
        parameter is large, the least-squares norm grows rapidly.
    }
    \label{fig:lcurve}
\end{figure}

Tikhonov regularization requires the observer to pick the value of $\varepsilon$. If $\varepsilon$
is too large then too much importance is placed on minimizing the norm of the solution and the
least-squares residuals will suffer. Conversely, if $\varepsilon$ is too small then the problem will
be poorly regularized and the resulting sky map may not represent the true sky. Picking the value of
$\varepsilon$ therefore requires understanding the trade-off between the two norms.

This trade-off can be analyzed quantitatively by trialing several values of $\varepsilon$, and
computing $\|\b v - \b B\b a\|^2$ and $\|\b a\|^2$ for each trial. An example is shown in
Figure~\ref{fig:lcurve}. The shape of this curve has a characteristic L-shape, and as a result this
type of plot is called an L-curve. The ideal value of $\varepsilon$ lies near the turning point of
the plot. At this point a small decrease in $\varepsilon$ will lead to an undesired rapid increase
in $\|\b a\|^2$, and a small increase in $\varepsilon$ will lead to an undesired rapid increase in
$\|\b v - \b B\b a\|^2$.

In practice, the L-curve should be used as a guide to estimate a reasonable value of $\varepsilon$.
However better results can often be obtained by tuning the value of $\varepsilon$. For instance,
increasing the value of $\varepsilon$ can improve the noise properties of the map by down-weighting
noisy modes. Decreasing the value of $\varepsilon$ can improve the resolution of the map by
up-weighting the contribution of longer baselines, which are likely fewer in number. In this
respect, choosing the value of $\varepsilon$ is analagous to picking the weighting scheme in
traditional imaging where robust weighting schemes can be tuned to similar effect \citep{briggs}.
For the OVRO-LWA we selected $\varepsilon = 0.01$ across all frequency channels. The distribution of
singular values of the transfer matrix with respect to $\sqrt{\varepsilon}$ is summarized in
Table~\ref{tab:summary}.

\subsubsection{Other Regularization Schemes}

The choice of applying Tikhonov regularization to $m$-mode analysis imaging is not unique. There
exists a plethora of alternative regularization schemes that could also be applied. Each
regularization scheme has its own advantages and disadvantages. For instance, Tikhonov
regularization is simple, independent of prior information, and sets unmeasured modes to zero (a
sensible expectation). We will now briefly discuss a few other alternatives.

The Moore-Penrose pseudoinverse (denoted with a superscript $\dagger$), is commonly applied to find
the minimum-norm linear-least squares solution to a set of linear equations. This can be used in
place of Tikhonov regularization as
\begin{equation}
    \b{\hat a}_\text{Moore-Penrose} = \b B^\dagger\b v\,.
\end{equation}
Much like Tikhonov regularization, the Moore-Penrose pseudoinverse sets components with small
singular values (below some user-defined threshold) to zero. Components with large singular values
(above the user-defined threshold) are included in the calculation at their full amplitude with no
down-weighting of modes near the threshold. The essential difference between using the Moore-Penrose
pseudoinverse and Tikhonov regularization is that the pseudoinverse defines a hard transition from
``on'' to ``off''. Modes are either set to zero or included in the map at their full amplitude. On
the other hand, Tikhonov regularization smoothly interpolates between these behaviors. Because of
this, Tikhonov regularization tends to produce better results in practical applications.

If the baselines have a noise covariance matrix $\b N \neq n\b I$ for some scalar $n$ (eg. the
interferometer is inhomogeneous), then the observer should minimize $(\b v-\b B\b a)^*\b N^{-1}(\b
v-\b B\b a) + \varepsilon\|\b a\|^2$. The matrix $\b N$ is used to weight the measurements such that
\begin{equation}
    \b{\hat a}_\text{min variance} = (\b B^*\b N^{-1}\b B + \varepsilon\b I)^{-1}
        \b B\b N^{-1}\b v\,.
\end{equation}

In the event that the observer has a prior map of the sky, $\|\b a - \b a_\text{prior}\|^2$ can be
used as the regularizing norm. This will use the prior map to fill-in missing information instead of
setting these modes to zero. In this case, the minimum is at
\begin{equation}
    \b{\hat a}_\text{with prior} = (\b B^*\b B + \varepsilon\b I)^{-1}
        (\b B^*(\b v - \b B\b a_\text{prior}))
        + \b a_\text{prior}\,.
\end{equation}
If instead the observer has a prior expectation on the covariance of the spherical harmonic
coefficients, Wiener filtering can also be used.  This technique is demonstrated for simulated
measurements by \citet{2016arXiv161203255B}.

Alternatively, we could opt to regularize the problem by enforcing smoothness in the sky maps. In
this case, the regularizing norm should be of the form $\|\nabla I(\hat r)\|^2$, where $\nabla I$ is
the gradient of the sky brightness in the direction $\hat r$. This is actually a generalization of
Tikhonov regularization where the objective function is $\|\b v-\b B\b a\|^2 + \varepsilon\|\b A\b
a\|^2$ for some matrix $\b A$. The minimum is at
\begin{equation}
    \b{\hat a}_\text{generalized} = (\b B^*\b B + \varepsilon\b A^*\b A)^{-1}\b B^*\b v\,.
\end{equation}

Finally, in many machine learning applications the $L_1$-norm\footnote{
    $\|\b a\|_1 = \sum_i |a_i|$
} is used in place of the usual $L_2$-norm in order to encourage sparsity in the reconstructed
signal. Applying this to $m$-mode analysis imaging would amount to minimizing $\|\b v-\b B\b a\|_2^2
+ \varepsilon\|\b a\|_1$. However, because we have decomposed the sky in terms of spherical
harmonics, the vector $\b a$ is not expected to be sparse. Consequently, the $L_1$-norm is generally
inappropriate for $m$-mode analysis imaging without an additional change-of-variables designed to
introduce sparsity.

\subsection{CLEAN}\label{sec:clean}

In traditional radio astronomy imaging, CLEAN \citep{1974A&AS...15..417H} is a physically motivated
algorithm that interpolates between measured visibilities on the $uv$-plane. In the absence of this
interpolation, gaps in the interferometer's $uv$-coverage are assumed to be zero, and -- in the
image plane -- sources are convolved with a point spread function (PSF) that is characteristic of
the $uv$-coverage.  Fundamentally, the interferometer's PSF is determined by which modes were
assumed to be zero in the initial imaging process.

In $m$-mode analysis imaging, we assumed modes were zero in two separate ways:
\begin{enumerate}
    \item We selected a set of spherical harmonic coefficients $a_{lm}$ to describe the sky
        brightness distribution. All modes with $l>l_\text{max}$ are neglected and assumed to be
        zero.
    \item Tikhonov regularization forces linear combinations of spherical harmonic coefficients with
        $\sigma_i \lesssim \sqrt{\varepsilon}$ towards zero.
\end{enumerate}
As a consequence, the final map of the sky is not assembled from a complete set of spherical
harmonics. Therefore, just as in traditional imaging, $m$-mode analysis imaging produces dirty maps
in which sources are convolved with a PSF.  This PSF can be improved by increasing the number and
variety of baselines, which increases the number of modes for which $\sigma_i \gg
\sqrt{\varepsilon}$.  Alternatively, by collecting more data, the signal-noise ratio of the measured
$m$-modes increases, which allows the observer to lower the value of $\varepsilon$ without
increasing the noise in the maps.  Finally, the CLEAN algorithm can be applied to interpolate some
of the missing information that was assumed to be zero.

\begin{figure*}[t]
    \begin{tabular}{c}
        \includegraphics[width=\textwidth]{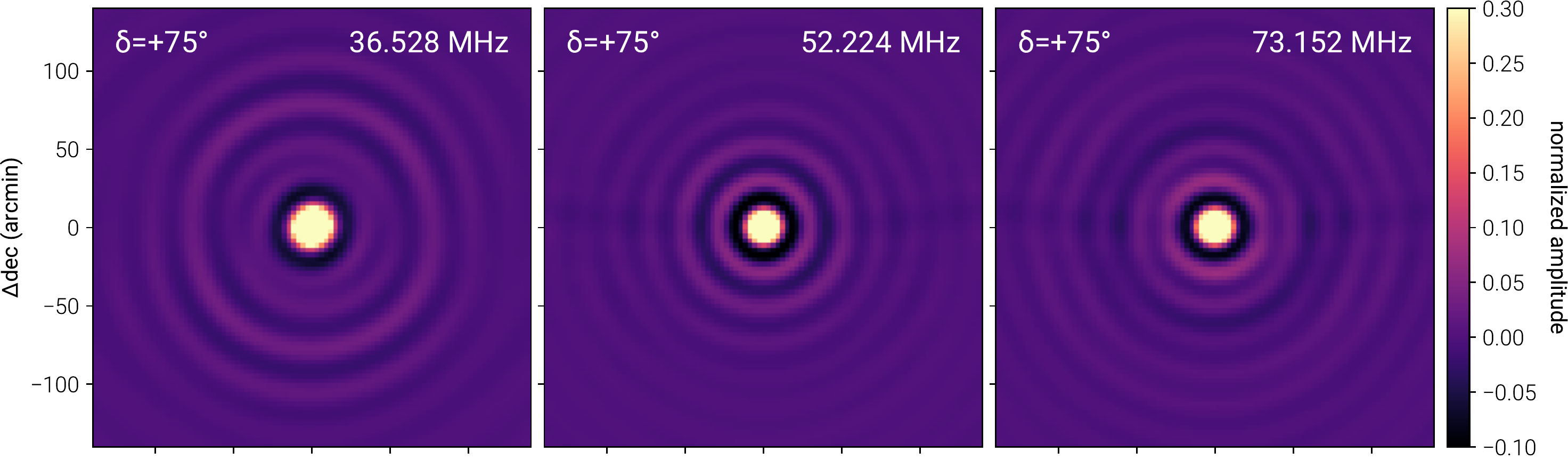} \\
        \includegraphics[width=\textwidth]{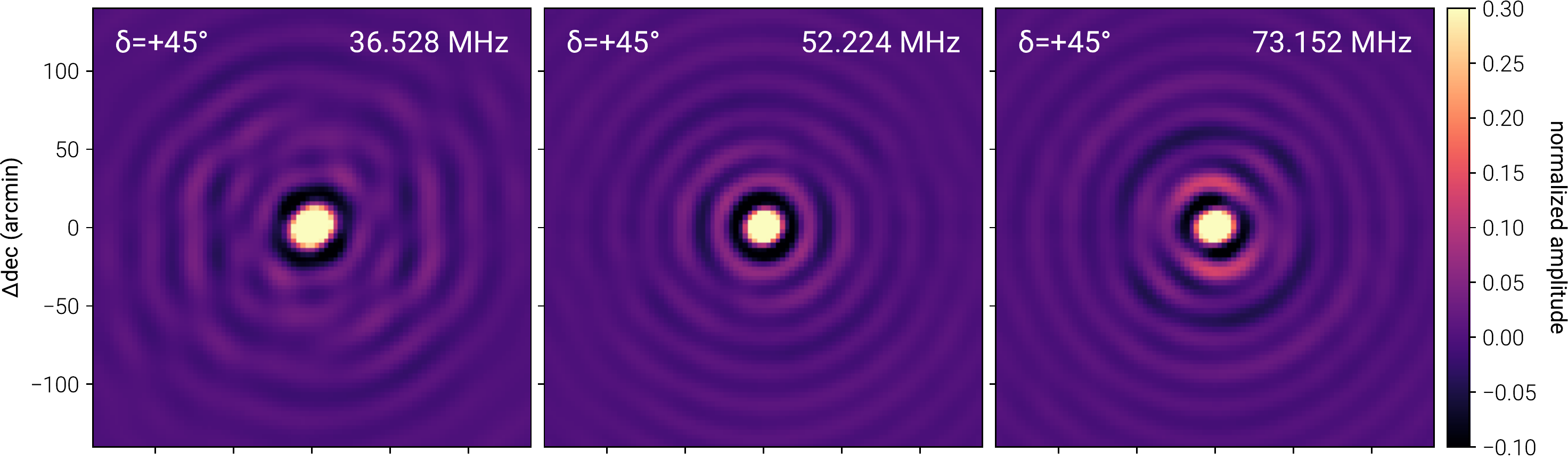} \\
        \includegraphics[width=\textwidth]{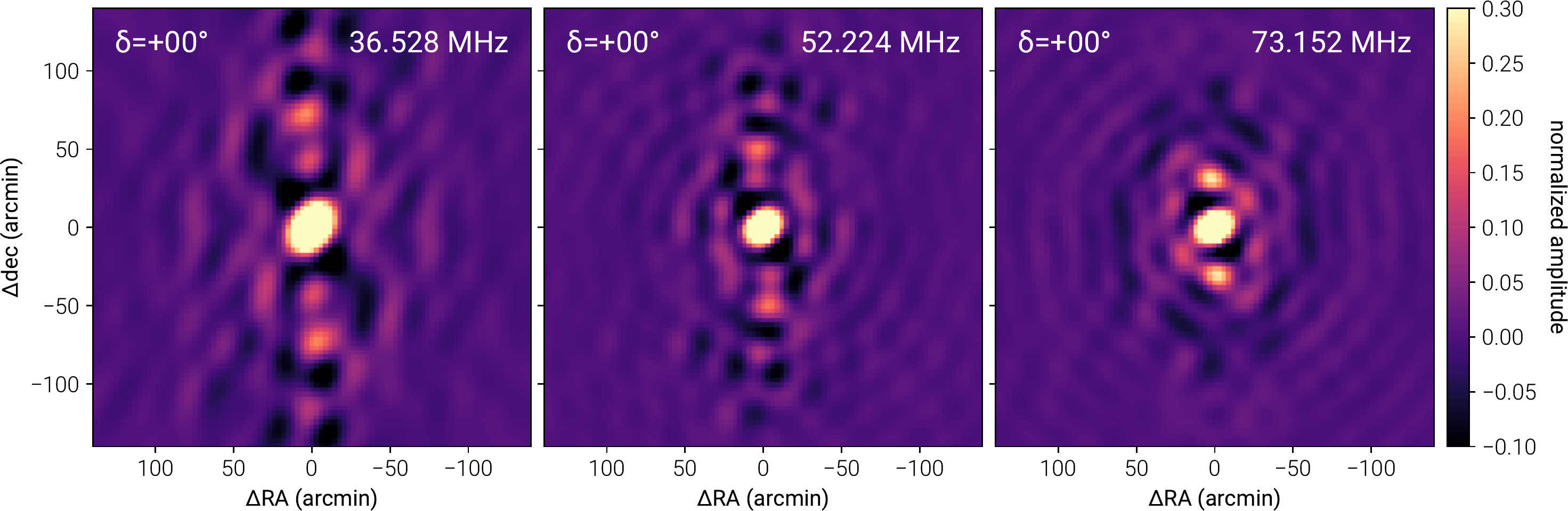} \\
    \end{tabular}
    \caption{
        The $m$-mode analysis imaging PSF at three declinations (top row: $\delta=+75^\circ$, middle
        row: $\delta=+45^\circ$, bottom row: $\delta=+0^\circ$) and three frequencies (left column:
        36.528~MHz, middle column: 52.224~MHz, right column: 73.152~MHz).  The PSF is computed by
        evaluating Equation~\ref{eq:psf}. Above 55~MHz, the angular extent of the PSF does not
        follow the expected scaling with frequency because the angular resolution is limited by the
        selection of $l_\text{max}=1000$. The FWHM at $\delta=+45^\circ$ is listed in
        Table~\ref{tab:summary}.
    }
    \label{fig:psf}
\end{figure*}

The PSF of a dirty $m$-mode analysis map may be computed with
\begin{equation}\label{eq:psf}
    \b a_\text{PSF}(\theta, \phi)
        = (\b B^*\b B + \varepsilon\b I)^{-1}\b B^*\b B\b a_\text{PS}(\theta, \phi)\,,
\end{equation}
where $\b a_\text{PSF}(\theta, \phi)$ is the vector of spherical harmonic coefficients representing
the PSF at the spherical coordinates $(\theta, \phi)$, and $\b a_\text{PS}(\theta, \phi)$ is the
vector of spherical harmonic coefficients for a point source at $(\theta, \phi)$ given by
\begin{equation}
    \b a_\text{PS}(\theta, \phi) = \begin{pmatrix}
        \vdots \\
        Y_{lm}^*(\theta, \phi) \\
        \vdots \\
    \end{pmatrix}
    = \begin{pmatrix}
        \vdots \\
        Y_{lm}^*(\theta, 0)\times e^{im\phi} \\
        \vdots \\
    \end{pmatrix} \,.
\end{equation}
In general, the PSF can be a function of the right ascension and declination. However, point sources
at the same declination take the same track through the sky and (barring any ionospheric effects)
will have the same PSF. The PSF is therefore only a function of the declination. For example,
sources at low elevations will tend to have an extended PSF along the north-south axis due to
baseline foreshortening. For the OVRO-LWA antenna configuration (Figure~\ref{fig:antenna-layout}),
example PSFs at three separate frequencies are shown in Figure~\ref{fig:psf}.  Adapting CLEAN for
$m$-mode analysis requires either pre-computing Equation~\ref{eq:psf} at a grid of declinations, or
a method for rapidly evaluating Equation~\ref{eq:psf} on the fly.

For an interferometer with more baselines than spherical harmonics used in the maps (eg. the
OVRO-LWA), $\b B^*\b B$ can be a much smaller matrix than the full transfer matrix $\b B$. Therefore
pre-computing $\b B^*\b B$ can allow the entire matrix to fit into memory on a single machine. This
greatly reduces the amount of disk I/O necessary for solving Equation~\ref{eq:psf}.

Additionally, we can pre-compute the Cholesky decomposition of $\b B^*\b B + \varepsilon\b I = \b
U^*\b U$, where $\b U$ is an upper-triangular matrix. Inverting an upper triangular matrix is an
$\mathcal{O}(N^2)$ operation (instead of $\mathcal{O}(N^3)$ for a general matrix inverse).\footnote{
    Instead of computing $\b A^{-1}$, we solve the linear equation $\b A\b x = \b b$ each time the
    matrix inverse is needed so as to avoid numerical instabilities.
}
Equation~\ref{eq:psf} can then be rapidly evaluated from right to left as:
\begin{equation}\label{eq:rapid-psf}
    \b a_\text{PSF} =
        \b U^{-1}\,\big(\b U^*\big)^{-1}\,\big(\b B^*\b B\big)\,\b a_\text{PS}\,.
\end{equation}
Furthermore, Equation~\ref{eq:rapid-psf} does not need to be separately evaluated for each CLEAN
component. Instead we can identify $N$ CLEAN components, accumulate $\b a_\text{PS}$ for each
component, and evaluate Equation~\ref{eq:rapid-psf} on the accumulation. This can greatly reduce the
number of times this equation needs to be evaluated, but care must be taken to ensure that the $N$
components are not so close together that sidelobes from one may interact with another.

Altogether the adaptation of CLEAN applied to the maps presented in this paper is summarized below.
\begin{algorithmic}[1]
    \Require{$\b a$ is the solution to Equation~\ref{eq:tikhonov-solution}}
    \Function{CLEAN}{$\b a$}
    \Let{$\b M$}{$\b B^*\b B$}
    \Let{$\b U$}{${\rm chol}(\b M + \varepsilon\b I)$} \Comment{Cholesky decomposition}
    \While{noise in map $>$ threshold}
    \State find $N$ pixels with the largest residual flux
    \Let{$\b x$}{$\sum_{i=1}^N \,(\text{pixel flux}) \times \b a_\text{PS}(\theta_i, \phi_i)$}
    \Let{$\b y$}{$\b U^{-1}\big(\b U^*\big)^{-1}\b M\b x$}
    \Let{$\b a$}{$\b a - (\text{loop gain})\times\b y$}
    \State record subtracted components
    \EndWhile
    \Let{$\b a$}{$\b a + (\text{restored components})$}
    \State \Return{$\b a$}
    \EndFunction
\end{algorithmic}

In summary, Tikhonov-regularized $m$-mode analysis imaging constructs a widefield synthesis image of
the sky from a complete earth rotation, and  with exact treatment of widefield-effects. This is
accomplished by solving a regularized block-diagonal matrix equation
(Equation~\ref{eq:tikhonov-solution}). The solution to this equation generates a map where
sources are convolved with a PSF characteristic of the interferometer (a function of the frequency,
antenna response, and baseline distribution with a full Earth rotation). The CLEAN algorithm is
adopted to deconvolve the PSF and produce the final sky maps.

\section{Observations}\label{sec:observations}

\subsection{The Owens Valley Long Wavelength Array}

\begin{figure}[t]
    \includegraphics[width=\columnwidth]{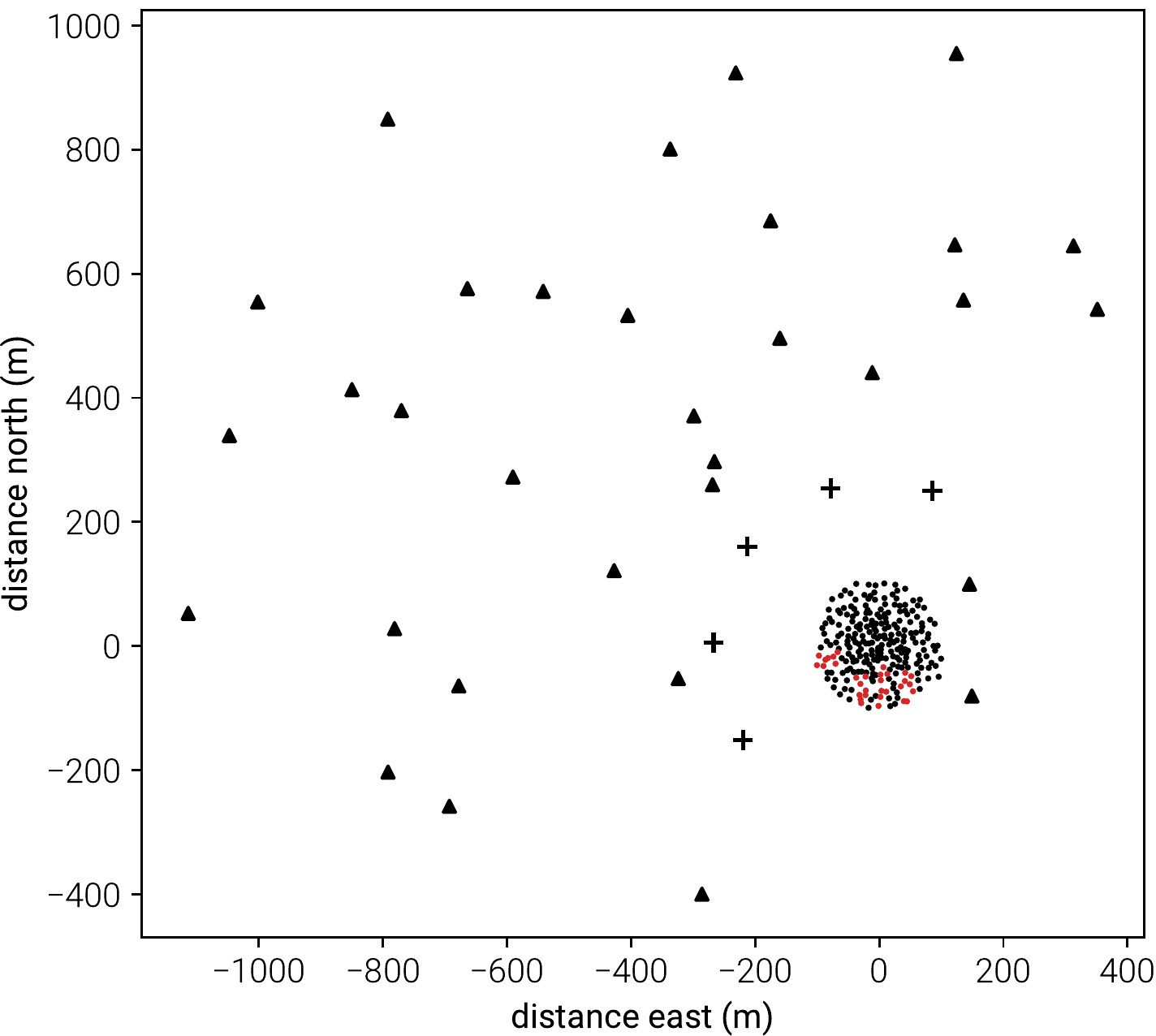}
    \caption{
        Antenna layout for the OVRO-LWA. Black dots correspond to antennas within the 200 m diameter
        core of the array. The 32 triangles are the expansion antennas built in early 2016 in order
        to increase the longest baseline to $\sim1.5$ km. The red dots are core antennas that are
        disconnected from the correlator in order to accommodate these antennas. The 5 crosses are
        antennas equipped with noise-switched front ends.
    }
    \label{fig:antenna-layout}
\end{figure}

The Owens Valley Long Wavelength Array (OVRO-LWA) is a 288-element interferometer located at the
Owens Valley Radio Observatory near Big Pine, California \citep{hallinan_2017}.  The OVRO-LWA is a
low-frequency instrument with instantaneous bandwidth covering 27~MHz to 85~MHz and with 24~kHz
channelization.  Each antenna stand hosts two perpendicular broadband dipoles so that there are
$288\times2$ signal paths in total. These signal paths feed into the 512-input LEDA correlator
\citep{2015JAI.....450003K}, which allows the OVRO-LWA to capture the entire visible hemisphere in a
single snapshot image.

The 288 antennas are arranged in a pseudo-random configuration optimized to minimize sidelobes in
snapshot imaging (see Figure~\ref{fig:antenna-layout}).  Of these 288 antennas, 251 are contained
within a 200 m diameter core, 32 are placed outside of the core in order to extend the maximum
baseline length to $\sim$1.5~km, and the final 5 antennas are equipped with noise-switched front
ends for calibrated total power measurements of the global sky brightness.  These antennas are used
as part of the LEDA experiment \citep{2017arXiv170909313P} to measure the global signal of 21-cm
absorption from the cosmic dawn. In the current configuration, 32 antennas (64 signal paths) from
the core are disconnected from the correlator in order to accommodate the 32 antennas on longer
baselines. A final stage of construction will involve 64 additional antennas installed on long
baselines out to a maximum length of 2.6~km.

The dataset used in this paper spans 28 consecutive hours beginning at 2017-02-17 12:00:00 UTC time.
During this time the OVRO-LWA operated as a zenith-pointing drift scanning interferometer.  The
correlator dump time was selected to be 13 seconds such that the correlator output evenly divides a
sidereal day. Due to the computational considerations presented in \S\ref{sec:mmode-analysis}, eight
24~kHz channels are selected for imaging from this dataset: 36.528~MHz, 41.760~MHz, 46.992~MHz,
52.224~MHz, 57.456~MHz, 62.688~MHz, 67.920~MHz, and 73.152~MHz.  These particular channels are
chosen due to their location at the exact center of instrumental subbands.

\subsection{Complex Gain Calibration}\label{sec:gaincal}

The complex gain calibration is responsible for correcting per-antenna amplitude and phase errors.
This is accomplished using a sky model and a variant of alternating least squares colloquially known
as ``Stefcal''
\citep{2008ISTSP...2..707M, 2014A&A...571A..97S}\footnote{
    The calibration routine is written in the Julia programming language
    \citep{doi:10.1137/141000671}, and is publicly available online
    (\url{https://github.com/mweastwood/TTCal.jl}) under an open source license (GPLv3 or any later
    version).
}.

Cyg~A and Cas~A are -- by an order of magnitude -- the brightest point-like radio sources in the
northern hemisphere at resolutions lower than 0.25$^\circ$. Therefore, the optimal time to solve for
the interferometer's gain calibration is when these sources are at high elevations.  The antenna
complex gains are measured from a 22 minute track of data when Cyg~A and Cas~A are at high
elevations. The gains measured in this way are then used to calibrate the entire 28 hour dataset.
The calibration sky model consists only of Cyg~A and Cas~A. The model flux of Cyg~A is set to the
\citet{1977A&A....61...99B} spectrum while the flux of Cas~A is measured from the data itself (using
a preliminary calibration solved for with a fiducial Cas~A spectrum).

Calibrating in this manner generates $\sim$ arcminute errors in the astrometry of the final sky maps
due to ionospheric refractive offsets during the time of calibration.  These residual errors in the
astrometry are corrected post-imaging by registering the images with respect to all VLSSr
\citep{2014MNRAS.440..327L} sources that are bright ($>30$~Jy with a consistent flux density
measured with the OVRO-LWA), and not too close to other bright sources (at least $1^\circ$
separation).

Temperature fluctuations of the analog electronics generate 0.1~dB sawtooth oscillations in the
analog gain. These oscillations occur with a variable 15 to 17 minute period associated with HVAC
cooling cycles within the electronics shelter that houses these electronics.  The amplitude of these
gain fluctuations is calibrated by smoothing the autocorrelation amplitudes on 45 minute timescales.
The ratio of the measured auto-correlation power to the smoothed auto-correlation power defines a
per-antenna amplitude correction that is then applied to the cross-correlations.  Additionally, the
ambient temperature at the front-end electronics (located in a box at the top of each dipole)
fluctuates diurnally, which will generate diurnal gain fluctuations. At this time, no correction is
made for these diurnal gain fluctuations.

\subsection{Primary Beam Measurements}\label{sec:beam}

\begin{figure*}[t]
    \includegraphics[width=\textwidth]{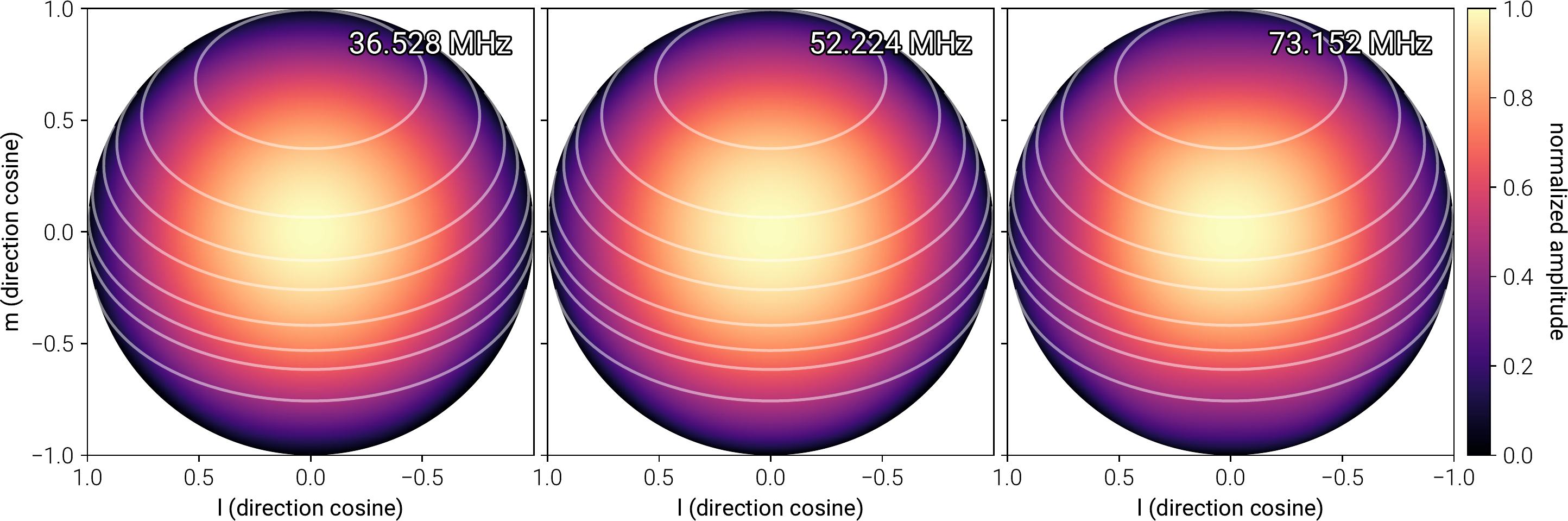}
    \caption{
        Empirical fits to the OVRO-LWA Stokes-I primary beam (the response of the $x$- and
        $y$-dipoles has been summed) at three frequencies: 36.528 MHz (left panel), 52.224 MHz
        (middle panel), and 73.152 MHz (right panel). The source tracks used to measure the beam
        model are overlaid. From north to south, these tracks correspond to Cas~A, Cyg~A, 3C~123,
        Tau~A, Vir~A, Her~A, 3C~353, and Hya~A.  The fitting process is described in
        \S\ref{sec:beam}, and residuals for Cyg~A and Cas~A are in Figure~\ref{fig:scintillation}.
    }
    \label{fig:beam}
\end{figure*}

In order to generate wide-field images of the sky, the response of the antenna to the sky must be
known. Drift-scanning interferometers like the OVRO-LWA can empirically measure their primary beam
under a mild set of symmetry assumptions \citep{2012AJ....143...53P}. The symmetry assumptions are
necessary to break the degeneracy between source flux and beam amplitude when the flux of the source
is unknown. In this work, we assume symmetries that are apparent in the antenna design, but
real-world defects and coupling with nearby antennas will contribute towards breaking these
symmetries at some level. In particular we assume that the $x$- and $y$-dipoles have the same
response to the sky after rotating one by 90$^\circ$, and that the beam is invariant under
north-south and east-west flips.

We measure the flux of several bright sources (Cyg~A, Cas~A, Tau~A, Vir~A, Her~A, Hya~A, 3C~123, and
3C~353) as they pass through the sky, and then fit a beam model composed of Zernike polynomials to
those flux measurements. We select the basis functions to have the desired symmetry ($Z_0^0$,
$Z_2^0$, $Z_4^0$, $Z_4^4$, $Z_6^0$, $Z_6^4$, $Z_8^0$, $Z_8^4$, and $Z_8^8$) and the beam amplitude
at zenith is constrained to be unity. See Figure~\ref{fig:beam} for an illustration of a fitted beam
model at several frequencies. This process is repeated for each frequency channel. Residuals for
Cyg~A and Cas~A can be seen in Figure~\ref{fig:scintillation}.

\subsection{Ionospheric Conditions}

\begin{figure*}[t]
    \includegraphics[width=\textwidth]{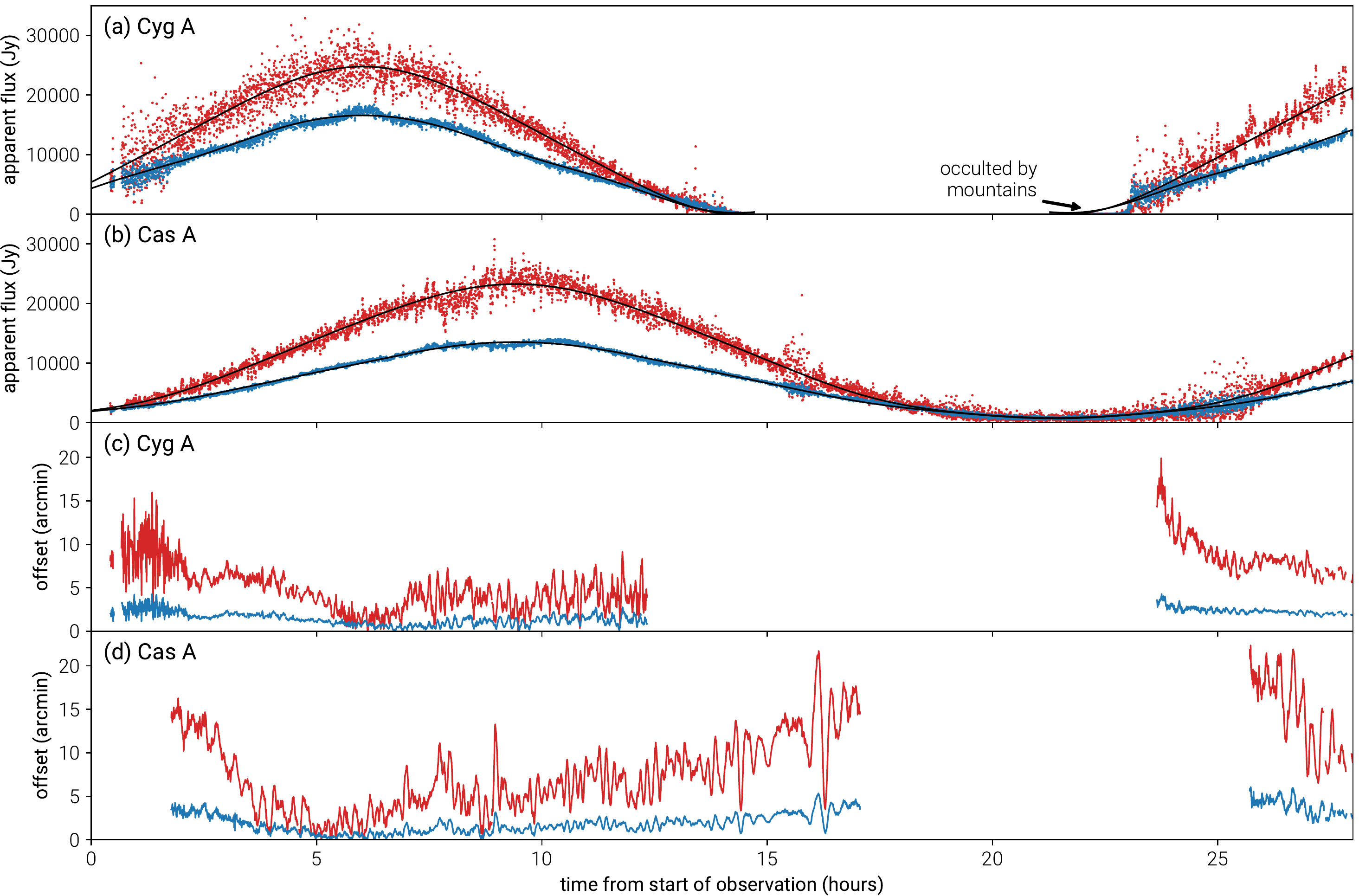}
    \caption{
        Panels (a) and (b) show the measured apparent flux of Cyg~A and Cas~A at 36.528 MHz (red
        points) and 73.152 MHz (blue points) as a function of time over the observing period. The
        solid black curves show the expected flux computed using the empirical beam model fits. The
        thermal noise contribution to each point is about 50 Jy.  Cyg~A is occulted by the White
        Mountains when it is low on the horizon to the east.
        Panels (c) and (d) These panels show the measured position offset of Cyg~A and Cas~A
        relative to their true astronomical positions at 36.528 MHz (red line) and 73.152 MHz (blue
        line).
    }
    \label{fig:scintillation}
\end{figure*}

\begin{figure}[t]
    \includegraphics[width=\columnwidth]{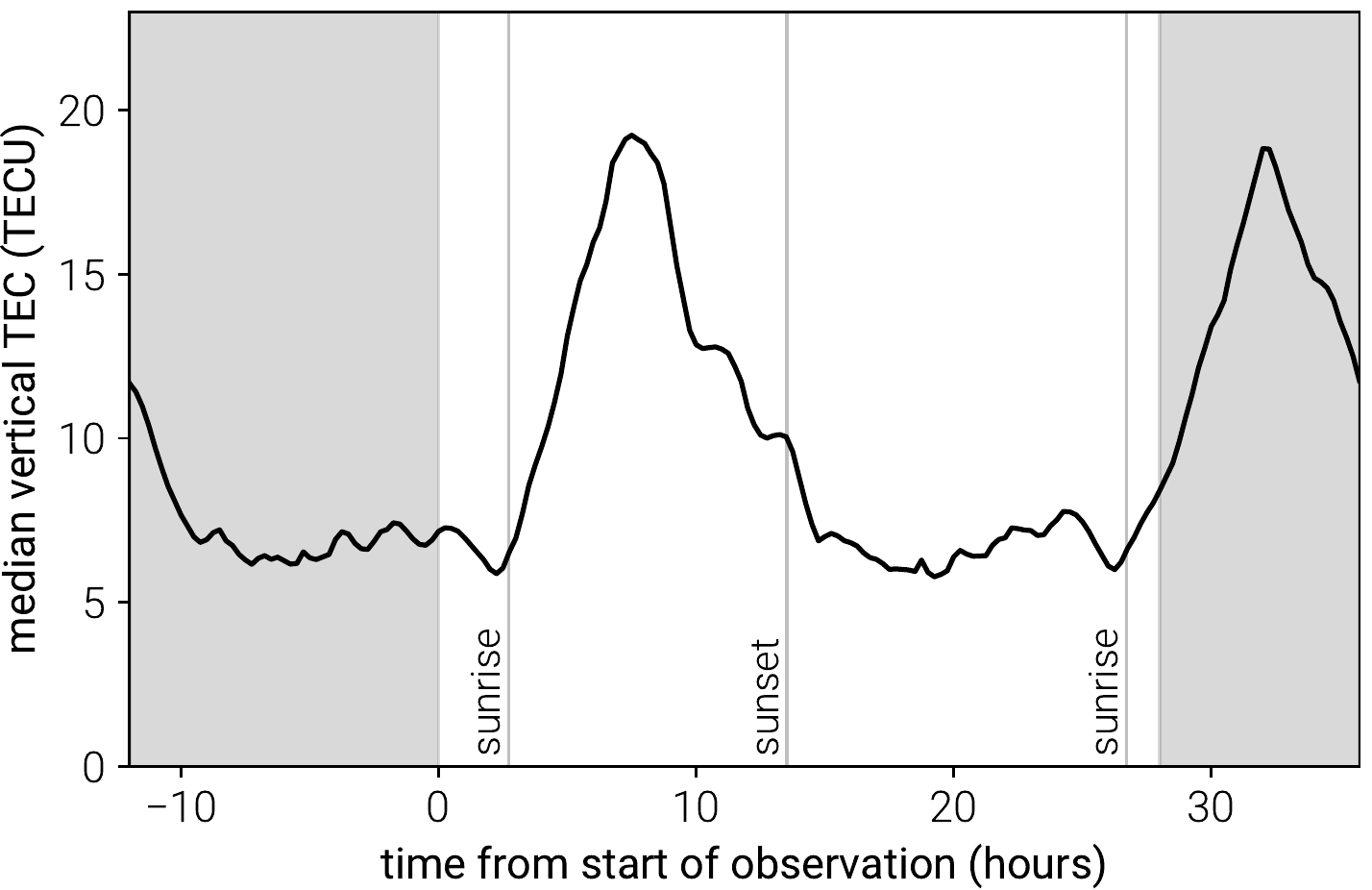}
    \caption{
        The median vertical TEC within 200 km of OVRO during the time of the observation. The gray
        shaded regions indicate times outside of the observing period. The gray vertical lines
        indicate sunrise and sunset (as labeled).
    }
    \label{fig:vtec}
\end{figure}

The geomagnetic conditions during this time were mild. The Disturbance storm time (Dst) index, which
measures the $z$-component of the interplanetary magnetic field, was
$>-30$ nT during the entirety of the observing period.\footnote{
    The Dst index was obtained from the World Data Center for Geomagnetism, Kyoto University
    (\url{http://swdcwww.kugi.kyoto-u.ac.jp/}). Accessed 2017-07-25.
}
Following the classification scheme of \citet{2008GMS...181.....K}, a weak geomagnetic storm has
$\text{Dst} < -30$ nT. Stronger geomagnetic storms have $\text{Dst} < -50$ nT.

Despite the mild conditions, low frequency interferometric observations are still affected by the
index of refraction in the ionosphere.  Figure~\ref{fig:vtec} shows the median vertical total
electron content (TEC) above OVRO measured from GPS \citep{1999JASTP..61.1205I}. The median is
computed over all GPS measurements within 200 km of the observatory. Over the observing period, the
TEC smoothly varies from 20 TECU at midday to 5 TECU during the night. However, this measurement is
only sensitive to large scale fluctuations in the ionosphere and does not capture small scale
fluctuations.

Small scale fluctuations are best characterized by source scintillation and refractive offsets.
Figure~\ref{fig:scintillation} shows the apparent flux and position offset of Cyg~A and Cas~A as a
function of time over the entire observing period. Both sources exhibit rapid scintillation on the
timescale of a single integration (13 seconds). For example, at 36.528 MHz, it is not unusual for
Cyg A to have measured flux variations of 50\% between adjacent 13 second integrations.  The
variance at 36.528~MHz compared with the variance at 73.152~MHz is consistent with an ionospheric
$\nu^{-2}$ origin. The measured position offset of each source is a measurement of the ionospheric
phase gradient across the array.  This varies on slower 10 minute timescales, with each source
refracting by as much as 20 arcmin (at 36.528 MHz) from their true astronomical positions as waves
in the ionosphere pass through the line of sight. At 74~MHz on the VLA, \citet{2007ApJS..172..686K}
observe $\sim 1$~arcmin refractive offsets during the night, and $\sim 4$~arcmin offsets during the
day on similar $\sim10$ minute timescales, which is consistent with what is seen here.  The impact
of these effects on the sky maps is simulated in \S\ref{sec:ionosphere}.

\subsection{Source Removal}\label{sec:source-removal}

\subsubsection{Cygnus A and Cassiopeia A}

Due to the rapid and large ionospheric fluctuations seen in Figure~\ref{fig:scintillation}, CLEAN
cannot be relied on to accurately deconvolve bright sources.  However, without removing bright
sources from the data, sidelobes from these sources will dominate the variance in the sky maps.  At
74~MHz Cyg~A is a 15,000~Jy source \citep{1977A&A....61...99B}. A conservative estimate for the
confusion limit at 74 MHz with a 15~arcmin beam is 1000~mJy \citep{2014MNRAS.440..327L}. Therefore,
we require that Cyg~A's sidelobes be at least $-45$~dB down from the main lobe to prevent Cyg~A's
sidelobes from dominating the variance in the image.

In order to account for propagation effects through the ionosphere, direction-dependent calibration
and peeling \citep{2008ISTSP...2..707M, 2015MNRAS.449.2668S} must be used.  Direction-dependent
calibration allows the per-antenna amplitude and phase towards a bright astronomical source to be
free parameters. In the dataset used in this paper, scintillation and refractive-offset events on
the timescale of a single integration (13 seconds) are observed (Figure~\ref{fig:scintillation}).
Therefore, the direction dependent calibration changes on these timescales, and we must solve for
one set of complex gains per source per integration.

The largest angular scale of Cas~A is $\sim$8~arcmin, and the largest angular scale of Cyg~A is
$\sim$2~arcmin. With a $\sim10$~arcmin resolution on its longest baselines at 73~MHz, the OVRO-LWA
marginally resolves both sources. A resolved source model is needed for both sources. We fit a
self-consistent resolved source model to each source. This is performed by minimizing the variance
within an aperture located on each source after peeling. By phasing up a large number of
integrations before imaging (at least 1 hour), it is possible to smear out the contribution of the
rest of the sky.  We then use a non-linear optimization routine (NLopt Sbplx) \citep{nlopt, sbplx}
to vary the parameters in a source model until the variance within the aperture is minimized. Cyg~A
is modeled with 2 Gaussian components, while Cas~A is modeled with 3~Gaussian components. Armed with
these  source models, both sources can be peeled from the dataset to the required dynamic range.

\subsubsection{Other Bright Sources}

Other bright sources -- namely Vir~A, Tau~A, Her~A, Hya~A, 3C~123, and 3C~353 -- are also removed
from the visibilities prior to imaging. Because these sources are much fainter than Cyg~A and Cas~A,
we do not need resolved source models to be able to remove these sources from the visibilities
without residual sidelobes contaminating the image.

However, the ionosphere will cause these sources to scintillate and refract. The position and flux
of each source is measured separately in each channel and integration. The sources are then
subtracted from the visibilities using the updated position and flux of the source. The brightest of
these sources (Vir~A and Tau~A) are peeled using a direction dependent calibration when they are at
high elevations.

\subsubsection{The Sun}

The Sun can be trivially removed from any map of the sky by constructing the map using only data
collected at night. A map of the entire sky can be obtained by using observations spaced 6 months
apart.  However the dataset used in this paper consists of 28 consecutive hours. Fortunately the Sun
was not active during this period, which could have greatly increased the difficulty involved in
subtracting the Sun.

We attempt to subtract the Sun from the dataset with the goal of suppressing its sidelobes.  The Sun
is well-resolved by the OVRO-LWA and hence a detailed source model is needed. In fact, the optical
depth $\tau=1$ surface of the Sun changes with frequency, and as a consequence a new model is needed
at each frequency. While we could fit a limited number of Gaussian components to Cyg~A and Cas~A,
this is insufficient for the Sun.  Additionally, while most astronomical sources at these
frequencies have negative spectral indices, the Sun has a positive spectral index. Therefore, more
care will need to be taken in subtracting the Sun at higher frequencies than at lower frequencies.

The strategy used for removing the Sun below 55 MHz involves fitting a shapelet
\citep{2003MNRAS.338...35R} model to the Sun and subtracting without the use of direction dependent
gains. The shapelet fitting is performed in the visibility space. Above 55 MHz a model is fit to the
Sun by minimizing the residuals after peeling (in the same way that models are obtained for Cyg~A
and Cas~A). The Sun is then peeled from each integration using direction dependent gains.

\subsection{Flux Scale}

\begin{figure*}[t]
    \includegraphics[width=\textwidth]{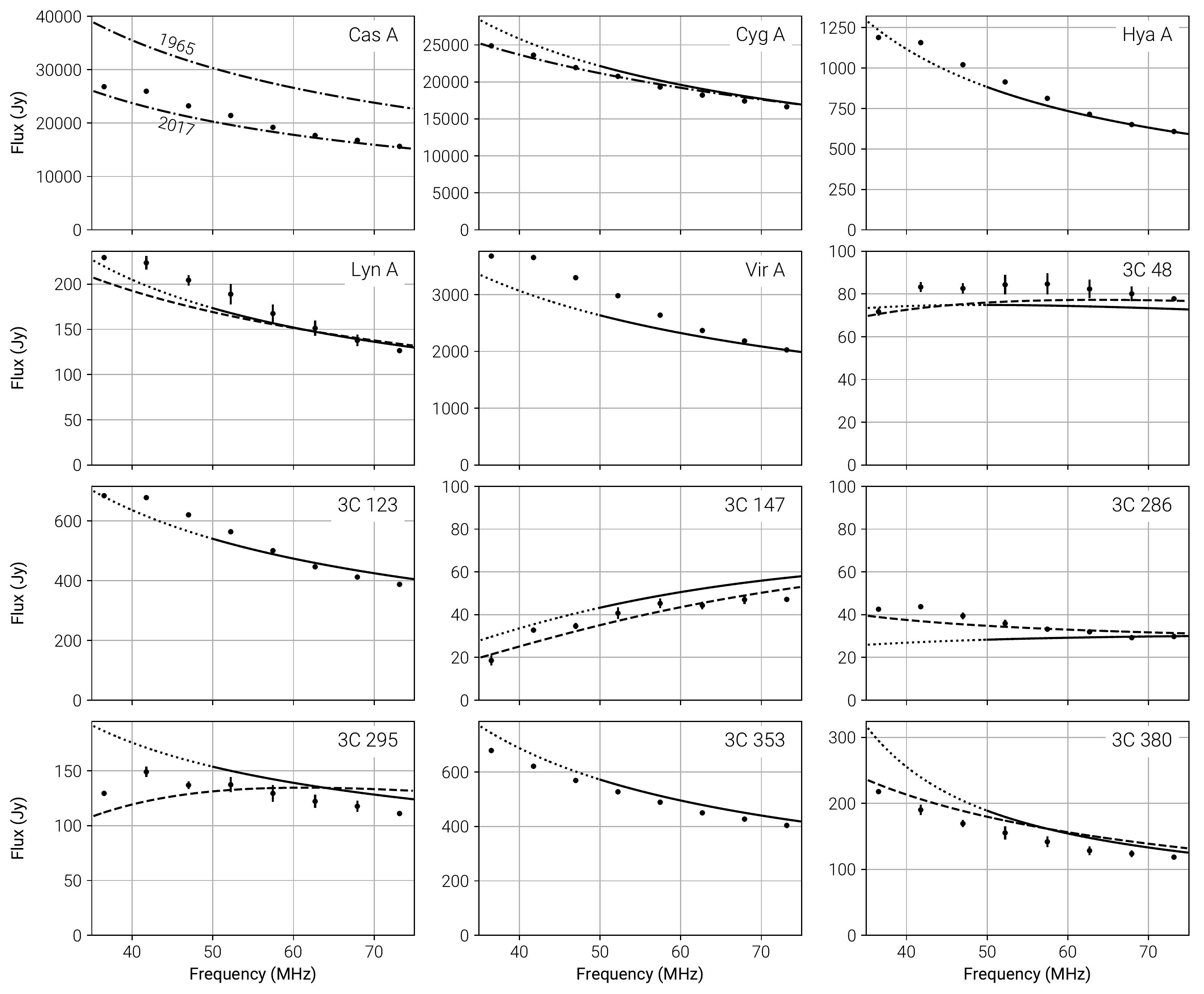}
    \caption{
        Measured fluxes (black points) of 11 sources plotted against the published spectra from
        \citet{2017ApJS..230....7P} (solid line above 50 MHz, dotted line below 50 MHz),
        \citet{2012MNRAS.423L..30S} (dashed line), and \citet{1977A&A....61...99B} (dot-dash line).
        Cas~A is compared against a spectrum assuming a secular decrease of 0.77\% per year
        \citep{2009AJ....138..838H}.
    }
    \label{fig:flux-scale}
\end{figure*}

The flux scale of the data was tied to the \citet{1977A&A....61...99B} spectrum of Cyg~A during gain
calibration. However, gain calibration is also a function of the beam model and the spectrum used
for Cas~A. Recent work by \citet{2012MNRAS.423L..30S} (hereafter SH12) using archival data from the
literature and \citet{2017ApJS..230....7P} (hereafter PB17) using the VLA has expanded the number of
low-frequency radio sources with calibrated flux measurements from one (Cyg~A) to eleven in total.
While the SH12 flux scale is valid between 30 MHz and 300 MHz, the PB17 flux scale is somewhat more
limited because the lowest frequency observations come from the VLA 4-band system. As a consequence,
the PB17 flux scale is not valid below 50 MHz.

Figure~\ref{fig:flux-scale} shows a comparison between flux measurements made using the all-sky maps
from this work, and spectra from the aforementioned flux scales. Generally the OVRO-LWA flux
measurements agree to between 5 and 10\% of the SH12 spectra. Below 50 MHz there can be substantial
departures with respect to the extrapolated PB17 spectra (for eg.  3C 286, 3C 295, and 3C 380), but
it is usually the case that we have much better agreement with the SH12 spectra. This indicates that
the PB17 spectra cannot be extrapolated below 50 MHz.

\section{Results}\label{sec:results}

\begin{figure*}[ht]
    \centering
    \begin{tabular}{c}
        \includegraphics[height=0.32\textheight]{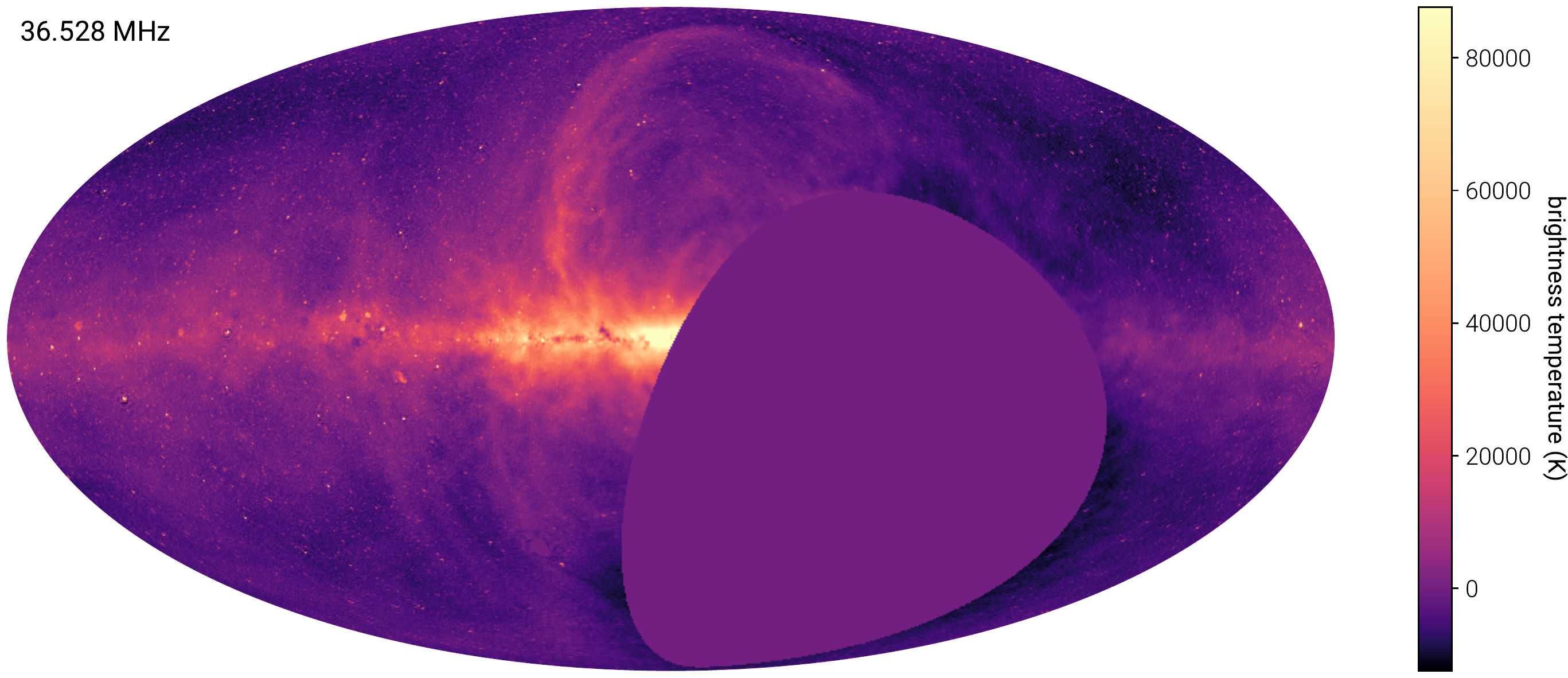} \\
        \includegraphics[height=0.32\textheight]{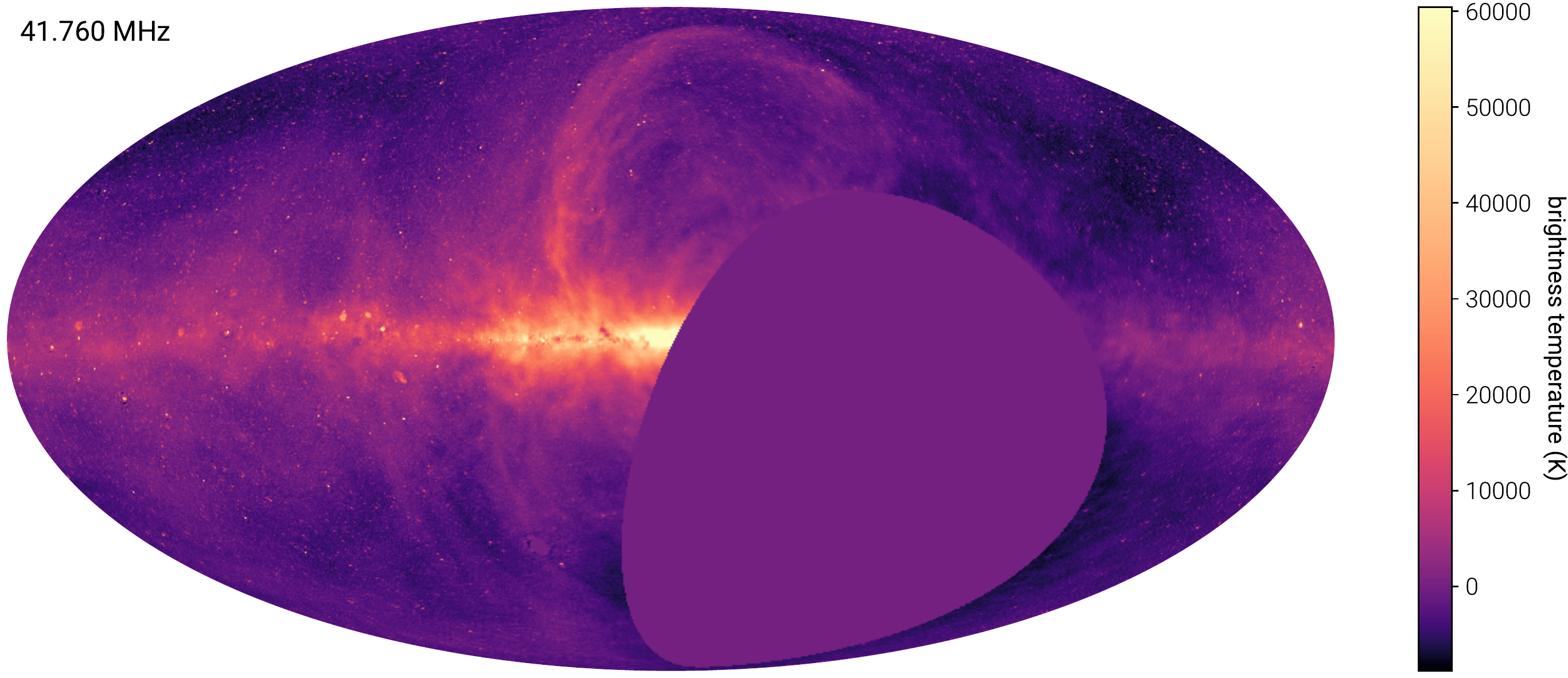} \\
    \end{tabular}
    \caption{
        These eight panels illustrate (with a Mollweide projection and logarithmic color scale) the
        eight full-sky maps generated with Tikhonov-regularized $m$-mode analysis imaging and the
        OVRO-LWA.  Each map covers the sky north of $\delta=-30^\circ$ with angular resolution of
        $\sim15$ arcmin. Eight bright sources have been removed from each map (Cyg~A, Cas~A, Vir~A,
        Tau~A, Her~A, Hya~A, 3C~123, and 3C~353). The small blank region near $l=+45.7^\circ$,
        $b=-47.9^\circ$ corresponds to the location of the Sun during the observation period.  A
        detailed summary of the properties of each map is given in Table~\ref{tab:summary}.
    }
    \label{fig:channel-maps}
\end{figure*}

\addtocounter{figure}{-1}
\begin{figure*}[p]
    \centering
    \begin{tabular}{c}
        \includegraphics[height=0.32\textheight]{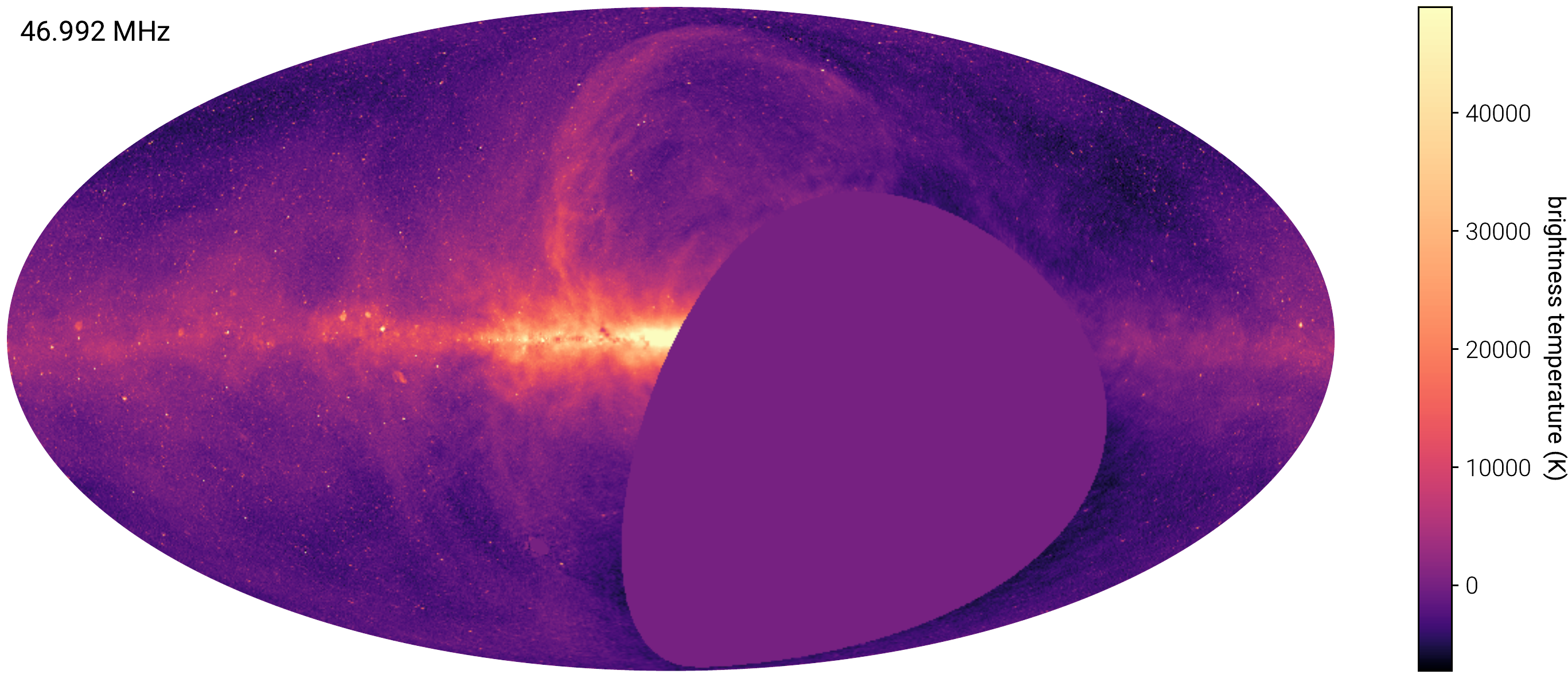} \\
        \includegraphics[height=0.32\textheight]{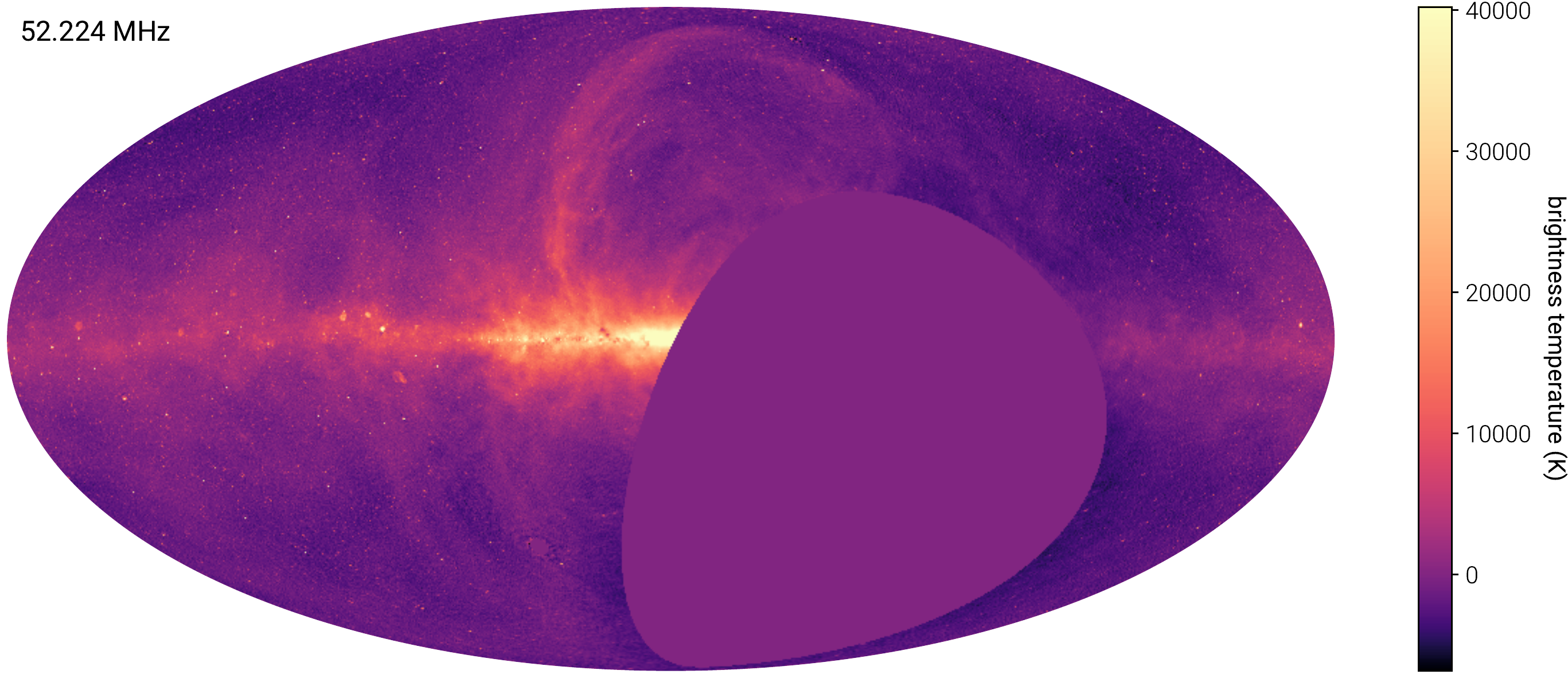} \\
        \includegraphics[height=0.32\textheight]{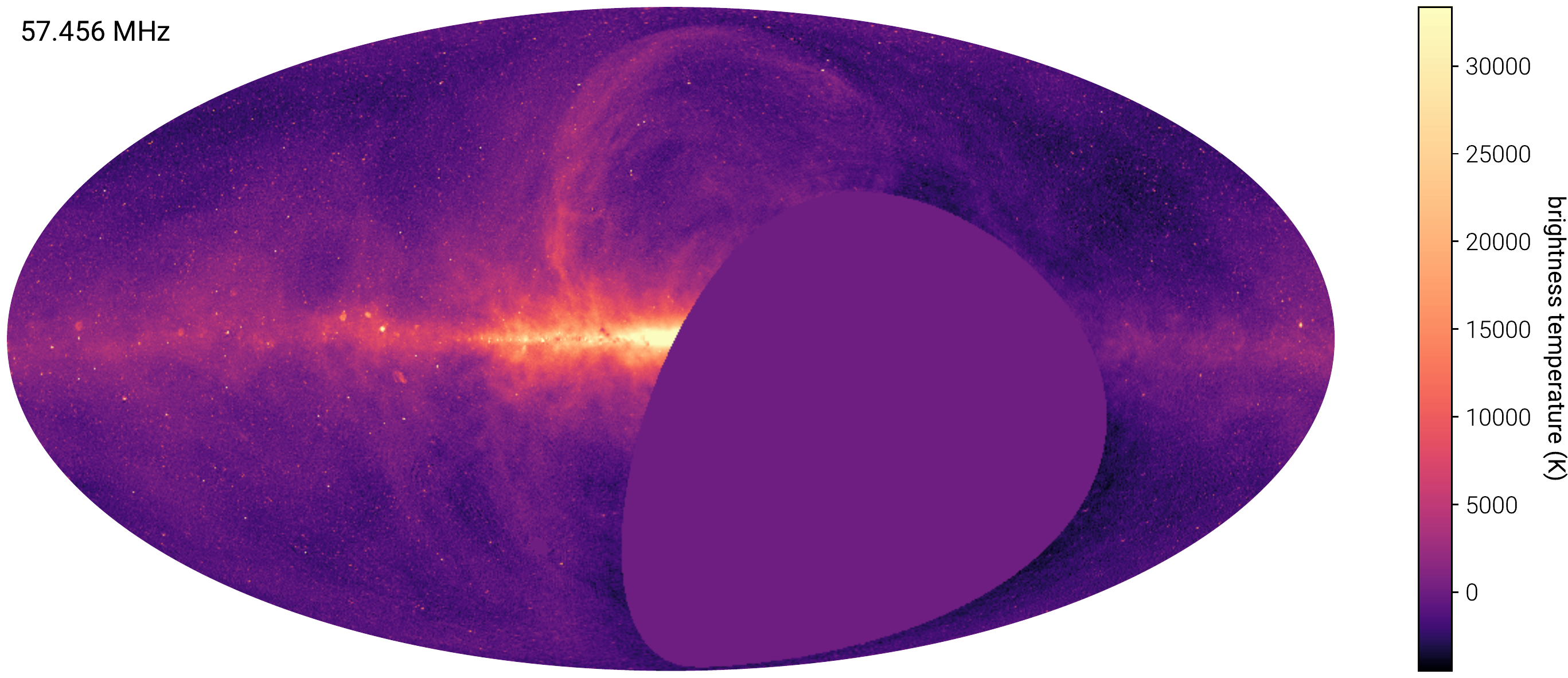} \\
    \end{tabular}
    \caption{
        continued
    }
\end{figure*}

\addtocounter{figure}{-1}
\begin{figure*}[p]
    \centering
    \begin{tabular}{c}
        \includegraphics[height=0.32\textheight]{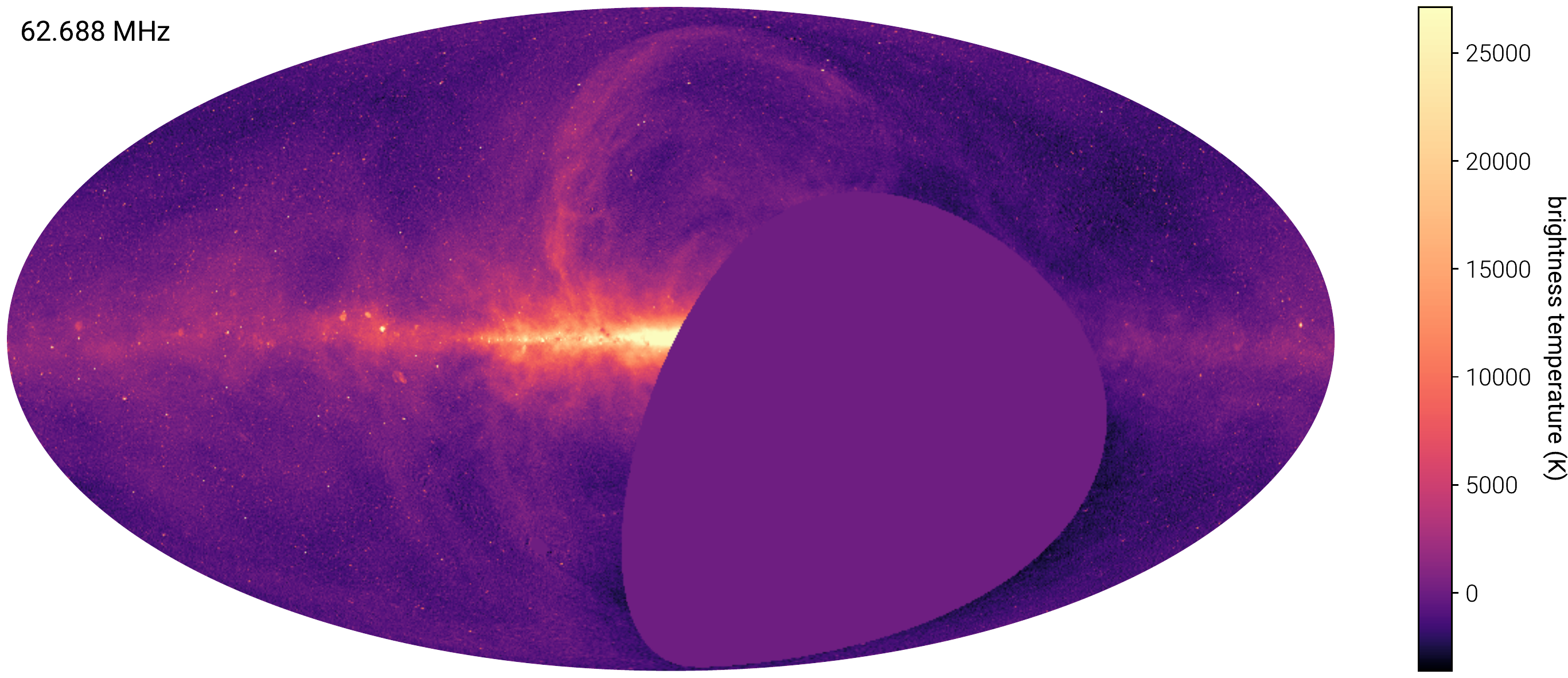} \\
        \includegraphics[height=0.32\textheight]{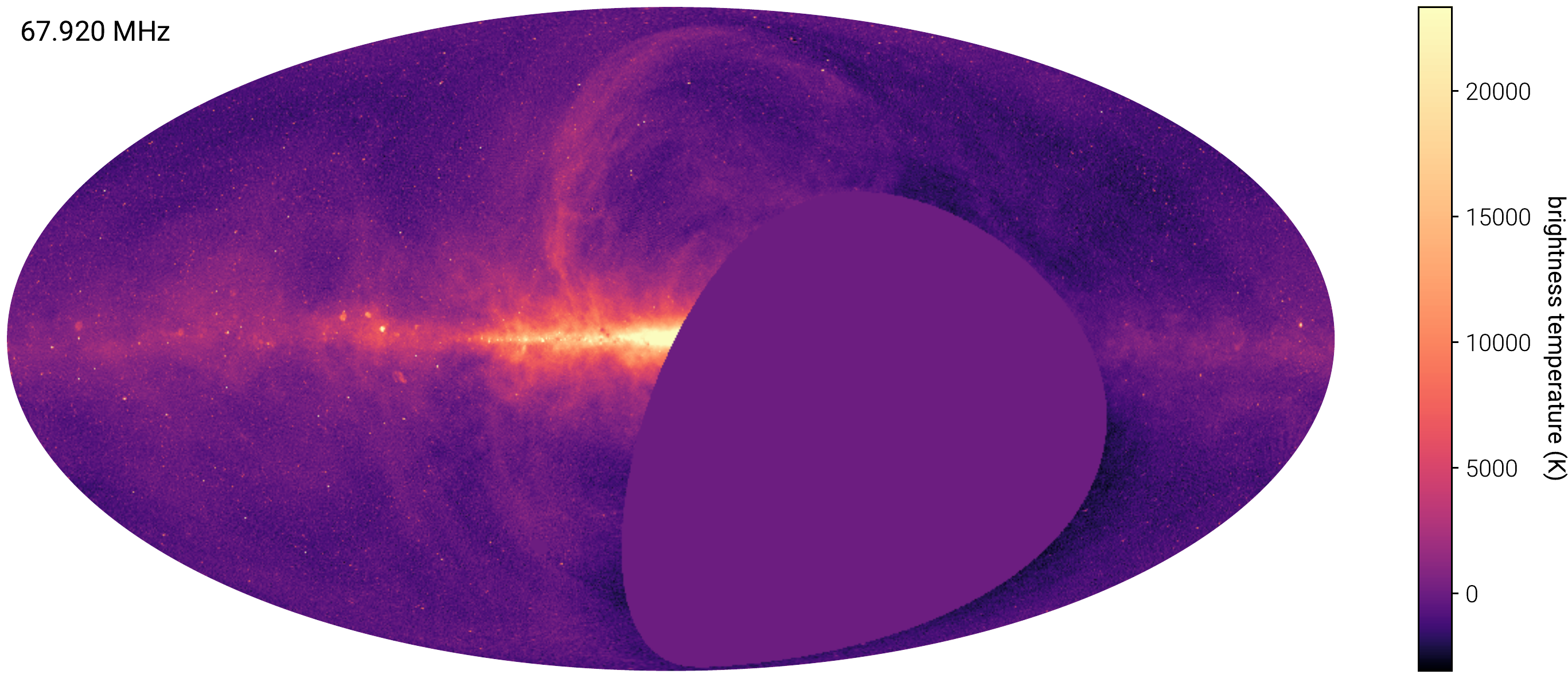} \\
        \includegraphics[height=0.32\textheight]{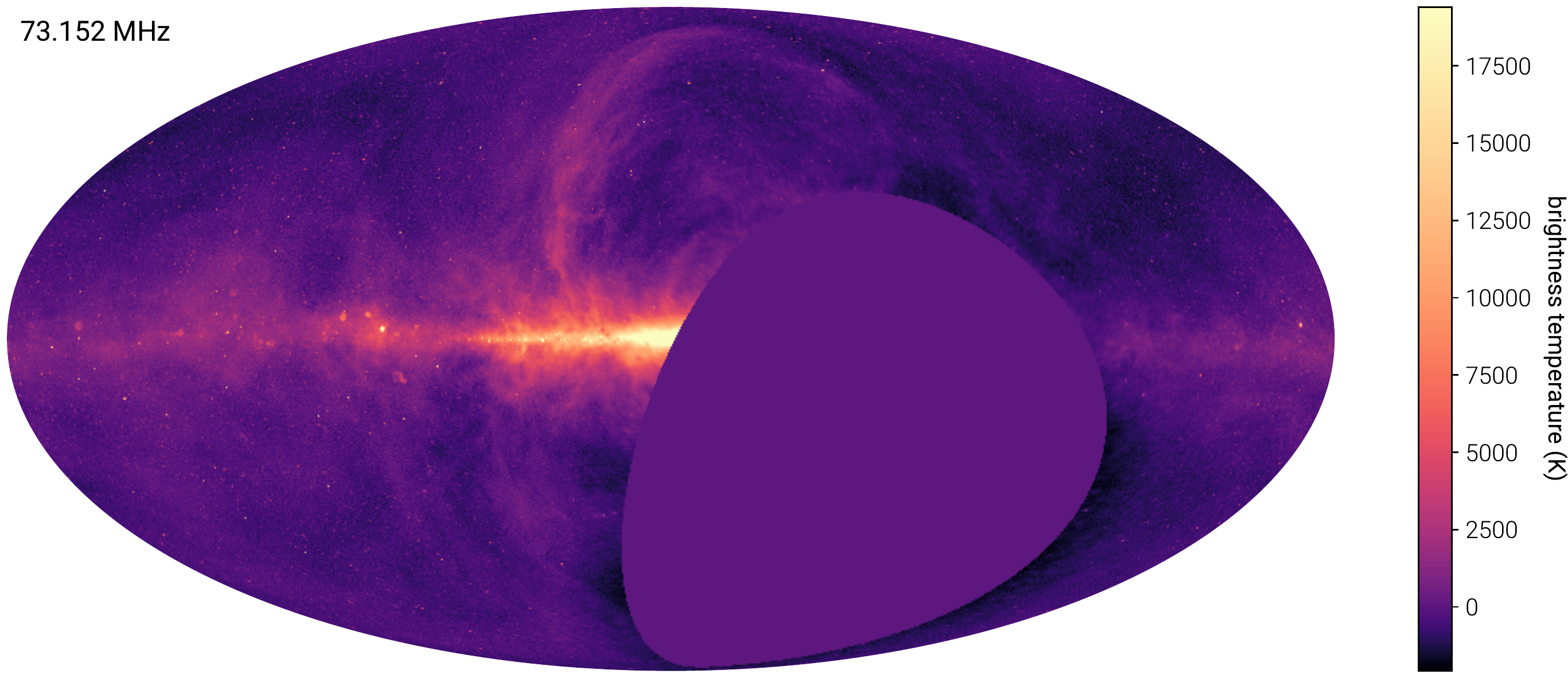} \\
    \end{tabular}
    \caption{
        continued
    }
\end{figure*}

\begin{figure*}[t]
    \centering
    \includegraphics[width=\textwidth]{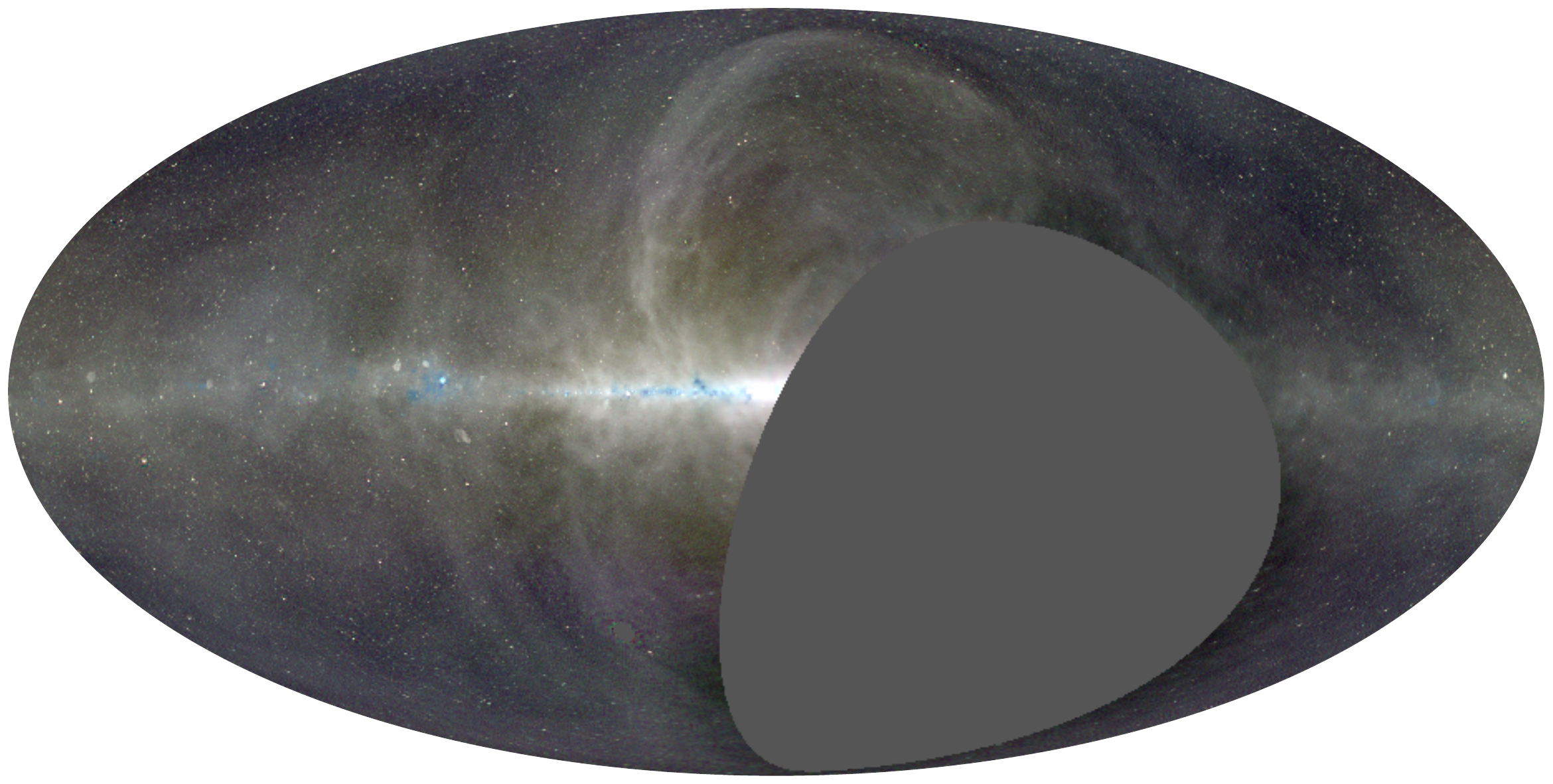}
    \caption{
        This Mollweide-projected map is constructed from three maps of the sky at 36.528~MHz (red),
        52.224~MHz (green), and 73.152~MHz (blue). The maps are scaled by $\nu^{2.5}$ before
        combining and the color scale is logarithmic (as in Figure~\ref{fig:channel-maps}).
        Therefore, regions with a spectral index of $-2.5$ will tend to appear white, and regions
        with a flatter spectral index will tend to appear blue.
    }
    \label{fig:three-color}
\end{figure*}

\begin{figure*}[t]
    \centering
    \includegraphics[width=\textwidth]{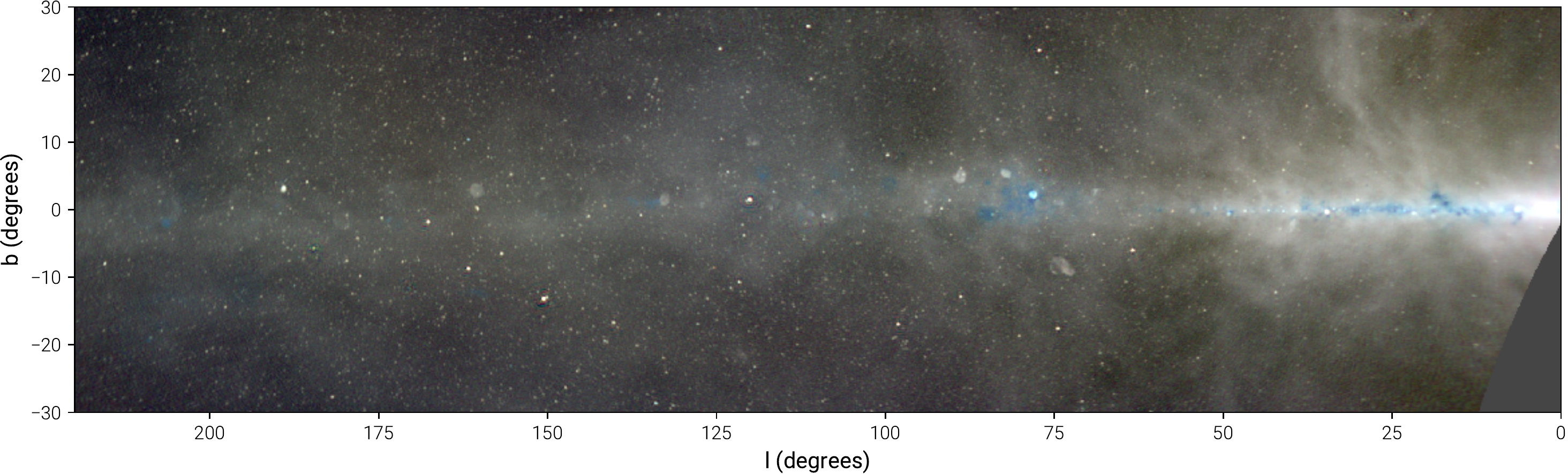}
    \caption{
        A cutout of the galactic plane from Figure~\ref{fig:three-color}.
    }
    \label{fig:galactic-plane-cutout}
\end{figure*}

\begin{figure*}[t]
    \centering
    \includegraphics[width=\textwidth]{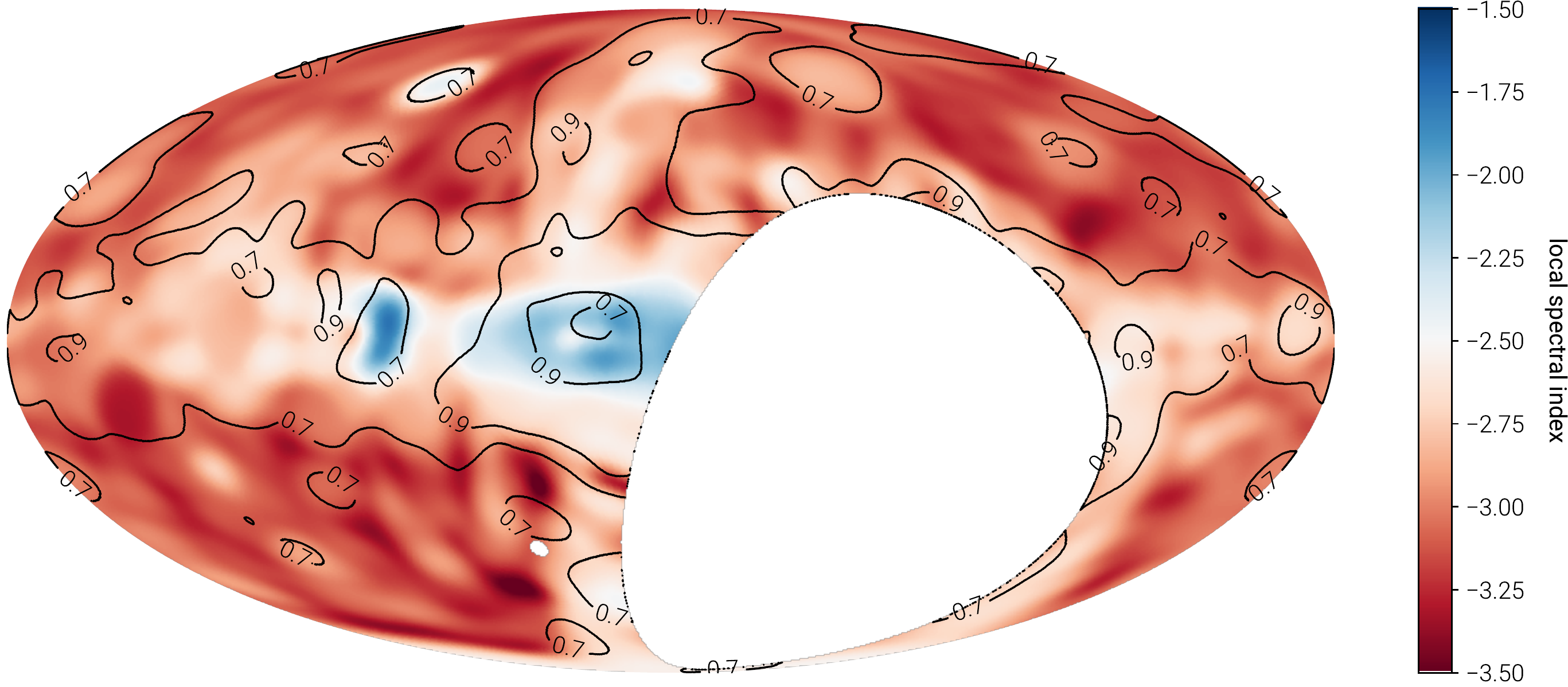}
    \caption{
        The local spectral index measured between the 36.528~MHz map and the 73.152~MHz map
        estimated by means of a local T-T plot. The color scale gives the spectral index where blue
        is flat spectrum and red is steep spectrum. The contours give the
        coefficient-of-determination ($R^2$) for the linear fit to the local T-T plot. If $R^2$ is
        low, the quality of the fit is low and the estimated spectral index is unreliable. This can
        be either due to insufficient dynamic range in the local T-T plot or due to multiple
        emission mechanisms operating in close proximity. Consequently $R^2$ tends to drop at higher
        galactic latitudes (due to dynamic range), and near \ion{H}{2} regions in the galactic
        plane (due to multiple emission mechanisms).
    }
    \label{fig:internal-spectral-index}
\end{figure*}

\begin{figure*}[t]
    \centering
    \includegraphics[height=0.32\textheight]{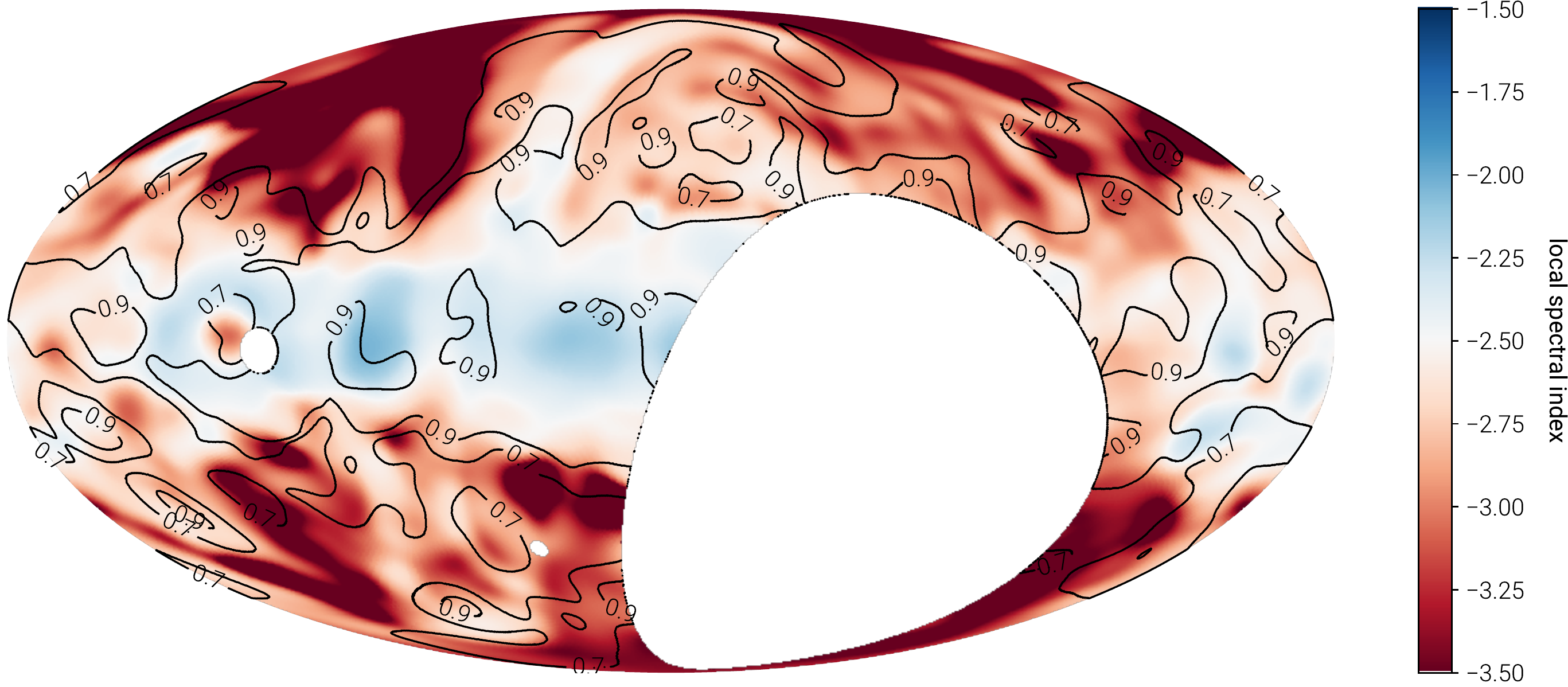}
    \caption{
        The local spectral index measured between the 73.152~MHz OVRO-LWA map and the reprocessed
        408~MHz Haslam map \citep{2015MNRAS.451.4311R}.  The color scale gives the spectral index
        where blue is flat spectrum and red is steep spectrum. The contours give the
        coefficient-of-determination ($R^2$) for the linear fit to the local T-T plot. See the
        caption of Figure~\ref{fig:internal-spectral-index} for more details about the
        coefficient-of-determination..
    }
    \label{fig:haslam-spectral-index}
\end{figure*}

\begin{table*}[t]
    \centering
    \begin{tabular}{cccccccccc}
        \hline
        \hline
        & \tbf{$\b\nu$}
        & \tbf{$\b\Delta\b\nu$}\footnote{
            Bandwidth used to construct the map. As described in the text, each map is constructed
            from a single frequency channel (24~kHz).
        }
        & \multicolumn3c{\tbf{FWHM}\footnote{
            The full-width half-maximum (FWHM) of the synthesized beam at the specified declination
            (major axis $\times$ minor axis).
        }}
        && \multicolumn2c{\tbf{Noise}\footnote{
            Measured with a jackknife and splitting the dataset into even- and odd-numbered
            integrations. This estimate therefore includes all noise sources that act on the
            timescale of a single 13 second integration (eg. thermal, ionospheric, etc.).
        }}
        & \tbf{Fraction of Modes}\footnote{
            Singular values of the transfer matrix compared with the value of the regularization
            parameter $\varepsilon$ used while solving Equation~\ref{eq:tikhonov-solution}. As
            discussed in the text, singular vectors with corresponding singular values $\sigma \ll
            \sqrt{\varepsilon}$ are set to zero by the Tikhonov regularization procedure.
        } \\
        \cline{4-6}
        \cline{8-9}
        \tbf{\#}
            & MHz & MHz
            & $\delta=0^\circ$ & $\delta=+45^\circ$ & $\delta=+75^\circ$
            && K & mJy/beam
            & with $\sigma>\sqrt{\varepsilon}$ \\
        \hline
        1 & 36.528 & 0.024 & $26.0'\times19.1'$ & $20.2'\times16.9'$ & $19.8'\times18.7'$ && 595. & 799. & 0.391 \\
        2 & 41.760 & 0.024 & $23.3'\times17.5'$ & $18.5'\times16.0'$ & $18.3'\times17.4'$ && 541. & 824. & 0.480 \\
        3 & 46.992 & 0.024 & $20.9'\times16.3'$ & $17.4'\times15.2'$ & $17.6'\times16.9'$ && 417. & 717. & 0.504 \\
        4 & 52.224 & 0.024 & $18.7'\times15.2'$ & $16.2'\times15.0'$ & $16.0'\times15.8'$ && 418. & 814. & 0.535 \\
        5 & 57.456 & 0.024 & $18.0'\times14.9'$ & $15.9'\times15.0'$ & $15.7'\times15.4'$ && 354. & 819. & 0.542 \\
        6 & 62.688 & 0.024 & $17.8'\times15.0'$ & $15.8'\times14.9'$ & $15.7'\times15.4'$ && 309. & 843. & 0.540 \\
        7 & 67.920 & 0.024 & $17.6'\times15.0'$ & $15.9'\times14.7'$ & $15.8'\times15.6'$ && 281. & 894. & 0.529 \\
        8 & 73.152 & 0.024 & $18.6'\times15.1'$ & $16.8'\times14.6'$ & $16.6'\times16.1'$ && 154. & 598. & 0.512 \\
        \hline \hline
    \end{tabular}
    \caption{A summary of the generated all-sky maps.}
    \label{tab:summary}
\end{table*}

We constructed eight sky maps using Tikhonov-regularized $m$-mode analysis imaging and CLEANing with
observations from the OVRO-LWA. Each map is individually shown in Figure~\ref{fig:channel-maps},
Figure~\ref{fig:three-color} is a three-color image constructed from the maps at 36.528~MHz,
52.224~MHz, and 73.152~MHz, and Figure~\ref{fig:galactic-plane-cutout} is a cutout of the galactic
plane. The maps cover the sky north of $\delta=-30^\circ$ with $\sim 15$ arcmin angular resolution.
The eight brightest northern hemisphere point sources are removed from each map (Cyg~A, Cas~A,
Vir~A, Tau~A, Her~A, Hya~A, 3C~123, and 3C~353) as described in \S\ref{sec:source-removal}, and
there is a small blank region near $l=+45.7^\circ$, $b=-47.9^\circ$ corresponding to the position of
the Sun during the observing window. The properties of each map -- including frequency, bandwidth,
angular resolution, and thermal noise -- are presented in Table~\ref{tab:summary}.

Each map from Figure~\ref{fig:channel-maps} will be made freely available online in Healpix format
\citep{2005ApJ...622..759G} on LAMBDA (\url{https://lambda.gsfc.nasa.gov/}).

Due to the considerations presented by \citet{2016ApJ...826..116V} and discussed in
\S\ref{sec:mmode-analysis}, each of these maps is monopole subtracted ($a_{00}=0$).  Furthermore, in
order to suppress sources of terrestrial interference, all spherical harmonics with $m=0$, or $m=1$
and $l>100$ are filtered from the map (where the spherical harmonics are defined in the J2017
coordinate system). As will be discussed in \S\ref{sec:rfi}, these spherical harmonics are
particularly susceptible to contamination by RFI and common-mode pickup. As a consequence,
astronomical emission that circles the J2017 north celestial pole (NCP) is filtered from the maps.
This filtering creates negative rings around the NCP at the declination of bright point sources.
These rings are naturally removed from the map during CLEANing as long as this filtering step is
included in the PSF calculation.

The noise in each map is empirically measured using jackknife resampling. The dataset is first split
into even- and odd-numbered integrations. These two groups are then imaged and CLEANed
independently, before being compared against the maps constructed from all of the available data
using the standard jackknife variance estimator. This estimate of the variance includes all sources
of error that operate on $\sim13$ second timescales (the integration time) such as thermal noise and
rapid ionospheric fluctuations, but does not account for more slowly varying effects (for example,
sidereal variation in the system temperature or day-night fluctuations in the ionosphere). These
noise calculations are summarized in Table~\ref{tab:summary}.  VLSSr source counts
\citep{2014MNRAS.440..327L} suggest that the confusion limit at 74 MHz and 15 arcmin angular
resolution is $\sim 1000\times(\nu/74\,{\rm MHz})^{-0.7}$ mJy.  Each channel map achieves thermal
noise $<900$ mJy and therefore each map is likely at or near the confusion limit.

In the absence of a zero-level correction, a pixel-by-pixel power law fit to the new maps is
impossible. In general this zero-level correction requires calibrated total power measurements that
will be included in future work.  Instead, temperature-temperature plots (T-T plots) can be used to
measure the spectral index independently of any zero-level corrections \citep{1962MNRAS.124..297T}.
This method relies on the assumption that all pixels in a given region are described by the same
power law. In that case, there exists a linear relationship between the brightness temperature at
frequency $\nu_1$ and frequency $\nu_2$. The slope of this best-fit line is a measure of the
spectral index between the two frequencies. T-T plots can fail to obtain a reliable measure of the
spectral index in two ways.  First, if there is not enough dynamic range in the emission region
there may be only a weak correlation between the brightness temperature at $\nu_1$ and $\nu_2$.
Second, if two emission mechanisms operate in close proximity (ie.  synchrotron and free-free), then
a single power-law interpretation of the emission in that region will be poor. Consequently,
spectral indices estimated from T-T plots can require careful interpretation.

In Figure~\ref{fig:internal-spectral-index}, the spectral index is locally estimated in each part of
the sky within a region $\sim10^\circ$ across by constructing local T-T plots between 36.528~MHz and
73.152~MHz. Contours of constant $R^2$ (the coefficient-of-determination) are overlaid. If $R^2\sim
1$, the spectral index is reliable because there is locally a strong linear correlation between
36.528~MHz and 73.152~MHz. However, if $R^2\ll 1$, the spectral index calculation is unreliable.
$R^2$ tends to drop in cold patches of the sky where there is not enough dynamic range to find a
strong correlation between the two frequencies. $R^2$ also tends to drop in the vicinity of
\ion{H}{2} regions in the galactic plane due to multiple emission mechanisms violating the
assumption of a single spectral index. Therefore, we should restrict our interpretation of
Figure~\ref{fig:internal-spectral-index} to the galactic plane and north galactic spur. In the
galactic plane, the synchrotron spectral index varies between $\sim-2.5$ and $-2.75$. In the
vicinity of \ion{H}{2}, regions the spectral index flattens significantly.  These \ion{H}{2} regions
can be seen with higher resolution in Figure~\ref{fig:galactic-plane-cutout}. In
Figure~\ref{fig:galactic-plane-cutout}, \ion{H}{2} regions appear as blue shadows along the galactic
plane due to the increasing impact of free-free absorption at lower frequencies.

In the literature, the spectral index at low frequencies is commonly computed with respect to the
Haslam 408~MHz map \citep{1981A&A...100..209H, 1982A&AS...47....1H}, which was reprocessed by
\citet{2015MNRAS.451.4311R} to remove artifacts associated with $1/f$ noise and bright sources.
Figure~\ref{fig:haslam-spectral-index} displays the spectral index computed between the 73.152~MHz
map and the reprocessed Haslam map. The spectral index was estimated by degrading the 73.152~MHz map
to the resolution of the Haslam map and constructing local T-T plots in every direction. The
coefficient-of-determination is overlaid as a contour plot, however because
$\log(408\,\text{MHz}/73.152\,\text{MHz}) > \log(73.152\,\text{MHz}/36.528\,\text{MHz})$, the
spectral indices presented in Figure~\ref{fig:haslam-spectral-index} tend to be more robust than
those presented in Figure~\ref{fig:internal-spectral-index}. This is reflected by the fact that
$R^2$ is larger, but the interpretation must still generally be restricted to the galactic plane.

\subsection{Comparisons with Other Sky Maps}\label{sec:compare}

\subsubsection{LWA1 Low Frequency Sky Survey}

\begin{figure*}[t]
    \centering
    \includegraphics[width=\textwidth]{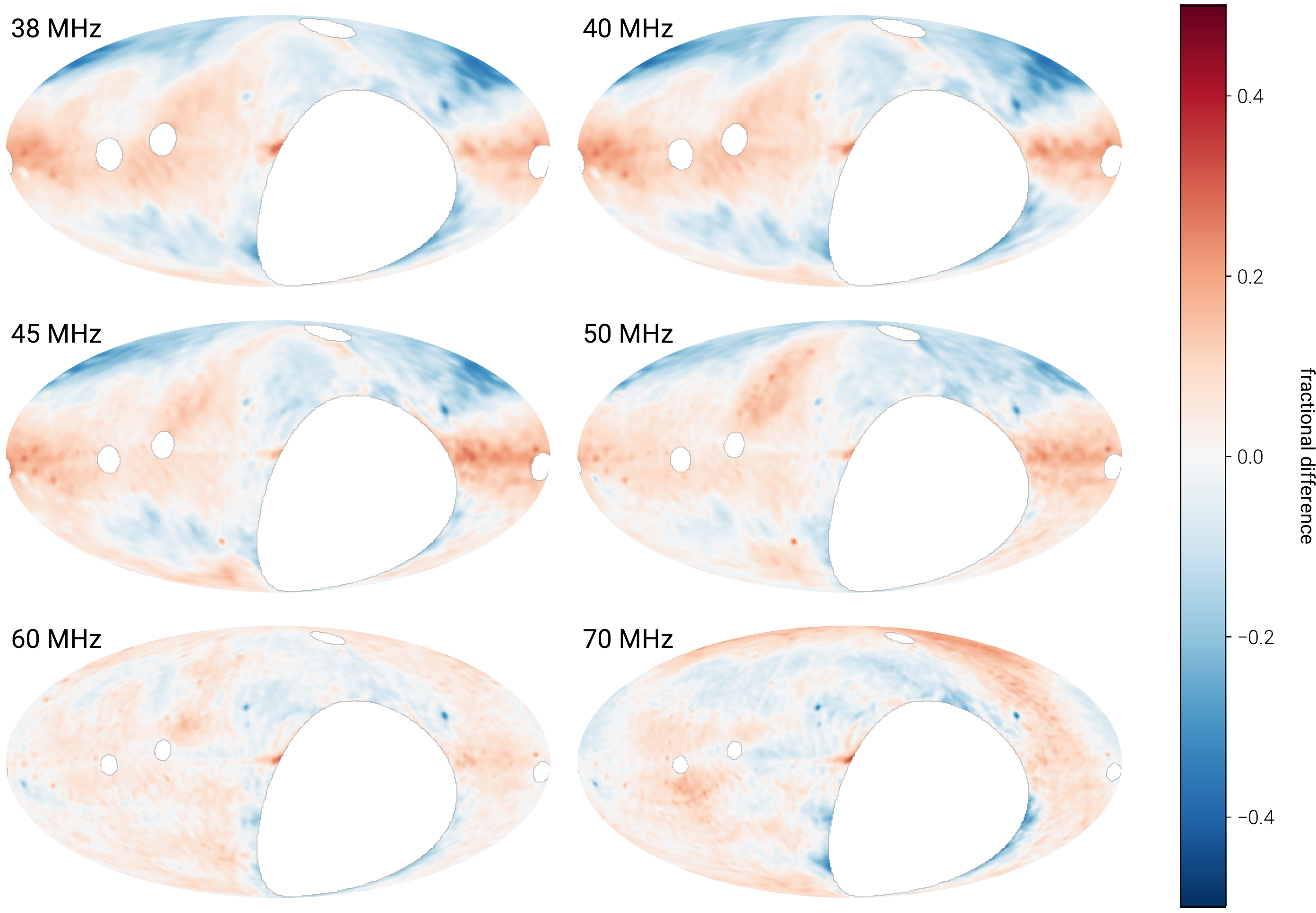}
    \caption{
        The fractional difference between maps from the LWA1 Low Frequency Sky Survey and the
        OVRO-LWA maps (Figure~\ref{fig:channel-maps}) after interpolating to the corresponding
        frequency and smoothing to the corresponding resolution. A positive value indicates regions
        where the OVRO-LWA map has more emission that the corresponding LWA1 Low Frequency Sky
        Survey map. Cas A, Cyg A, Vir A, and Tau A are masked due to the fact that they are
        subtracted from the OVRO-LWA maps.
    }
    \label{fig:lwa1-comparison}
\end{figure*}

The LWA1 Low Frequency Sky Survey (LLFSS) \citep{2017MNRAS.469.4537D} produced nine maps of the sky
between 35~MHz and 80~MHz. Six of these maps are interior to the frequency range spanned by this
work.  A direct comparison with these LLFSS maps can be seen in Figure~\ref{fig:lwa1-comparison}. In
this figure, the LLFSS maps are filtered to remove the monopole and all modes with $m=0$. The
OVRO-LWA maps are interpolated in frequency and blurred to match the angular resolution of the
corresponding LLFSS map. This comparison uncovered a $\sim45$~arcmin rotation in the LLFSS maps
about the NCP, which is corrected in Figure~\ref{fig:lwa1-comparison}.  At 60~MHz, the agreement is
generally better than 10\%. However at lower frequencies the agreement deteriorates to about 20\%.
Typically the OVRO-LWA maps have excess emission in the galactic plane and a deficit of emission off
the galactic plane relative to the LLFSS.

The LLFSS incorporates calibrated total power radiometry to estimate the missing flux from short
spacings. As a result \citet{2017MNRAS.469.4537D}, report per-pixel spectral-indices from combining
all nine sky maps. Care must be taken in comparing these spectral indices with
Figure~\ref{fig:internal-spectral-index} because they are susceptible to different systematic
errors. Both calculations are sensitive to mistakes in the antenna primary beam, but the LLFSS
spectral indices are additionally sensitive to errors in the zero-level. We will restrict the
comparison to the galactic plane where the spectral indices are likely to be the most reliable.
Towards the galactic center both surveys agree that the spectral index is very flat ($>-2.2$) due to
the influence of free-free absorption.  However, at galactic latitudes $\sim 180^\circ$ this work
suggests that the spectral index varies between -2.5 and -2.75, while the LLFSS reports
substantially flatter indices in the range -2.3 to -2.2. In this region $0.7 < R^2 < 0.9$ for the
OVRO-LWA, so this could be an artifact of the comparatively weak correlation between the brightness
at 36.528~MHz and 73.152~MHz, which tends to bias the spectral index towards $-\infty$.

The LLFSS also computes spectral indices with respect to the Haslam 408~MHz map. These spectral
indices are subject to the same caveats and systematic errors as before. However, in general the
qualitative agreement with Figure~\ref{fig:haslam-spectral-index} is better, potentially due to the
increased robustness associated with estimating spectral indices with a larger fractional bandwidth.

\subsubsection{Guzm\'{a}n 45 MHz Map}

\begin{figure*}[t]
    \centering
    \includegraphics[height=0.32\textheight]{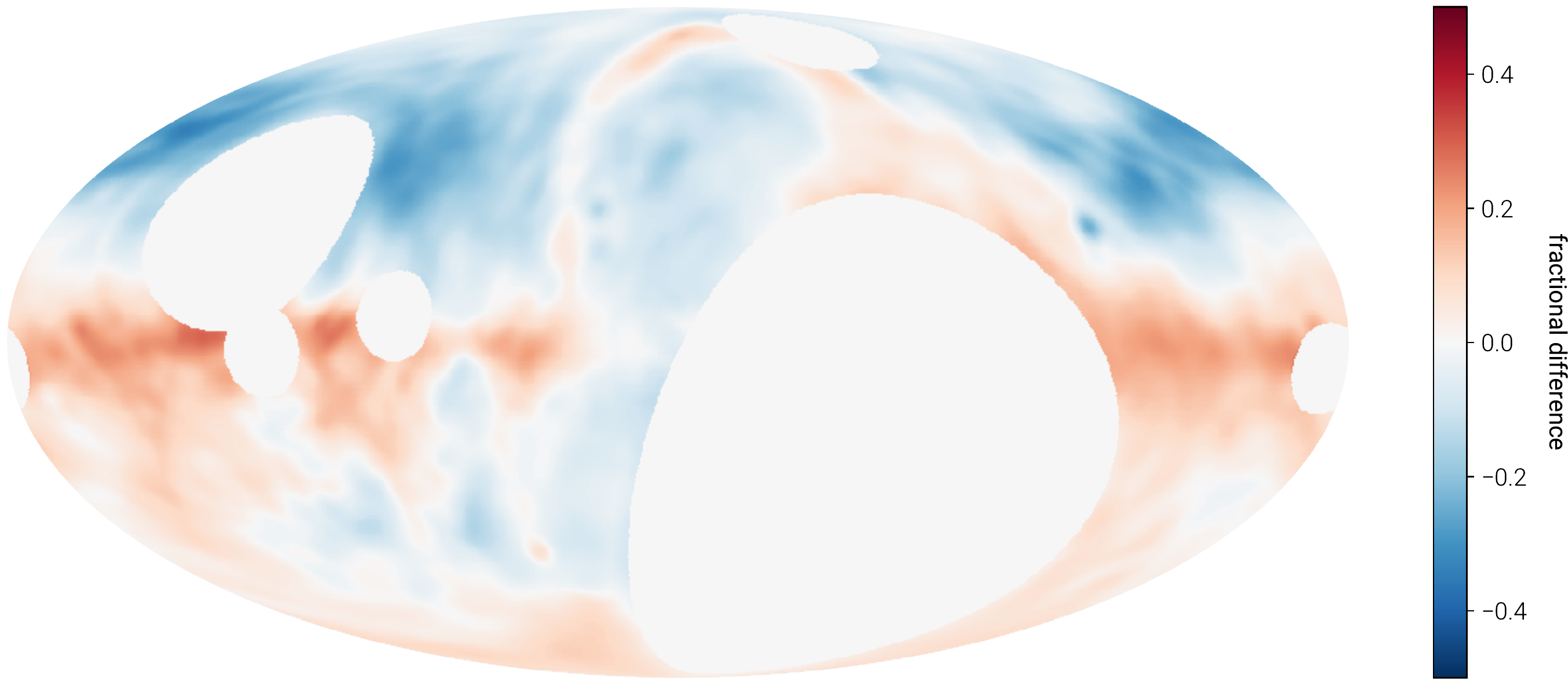}
    \caption{
        This Mollweide projected map compares the fractional difference between the Guzm\'{a}n 45
        MHz map, and the OVRO-LWA maps (Figure~\ref{fig:channel-maps}) interpolated to 45~MHz
        (degraded to $5^\circ$ resolution). A positive value indicates regions where the OVRO-LWA
        map has more emission that the Guzm\'{a}n map, and a negative value indicates regions where
        the Guzm\'{a}n map has more emission than the OVRO-LWA map. Cas A, Cyg A, Vir A, and Tau A
        are masked due to the fact that they are subtracted from the OVRO-LWA maps but not the
        Guzm\'{a}n map.
    }
    \label{fig:guzman-comparison}
\end{figure*}

The Guzm\'{a}n 45 MHz map \citep{2011A&A...525A.138G} is compiled from a southern hemisphere survey
\citep{1997A&AS..124..315A} and a northern hemisphere survey \citep{1999A&AS..140..145M} with a
small gap around the NCP. In this work, the zero-level is set by comparing against published
low-frequency measurements in six different directions.

A direct comparison between the OVRO-LWA maps interpolated to 45 MHz and the Guzm\'{a}n 45 MHz map
can be seen in Figure~\ref{fig:guzman-comparison}. In order to make this comparison, the OVRO-LWA
map was degraded to a 5$^\circ$ resolution by convolving with a Gaussian kernel, and the Guzm\'{a}n
map has had spherical harmonics with $m=0$ discarded in order to make it consistent with the maps
presented in this paper. This figure shows a $\sim20\%$ excess of emission in the galactic plane
that is consistent with the discrepancy observed between the LLFSS and the Guzm\'{a}n map.  However,
while the LLFSS has an excess of emission near the north galactic pole, no such excess is observed
in this work. Instead there is a 10\% excess of emission near the south galactic pole. Elsewhere off
the plane of the galaxy the discrepancy can be as much as $-20\%$.

\citet{2011A&A...525A.138G} compute the spectral index between their 45~MHz map and the 408~MHz
Haslam map. Along the galactic plane the spectral index varies between -2.2 (in the vicinity of
\ion{H}{2} regions), and -2.5 (at galactic longitudes $\sim 180^\circ$). The north galactic spur has
a spectral index of -2.5. This is generally consistent with the results presented in
Figure~\ref{fig:haslam-spectral-index}.

\section{Error Analysis}\label{sec:error}

\subsection{The Ionosphere}\label{sec:ionosphere}

\begin{figure*}[t]
    \centering
    \begin{tabular}{cc}
        \includegraphics[width=\columnwidth]{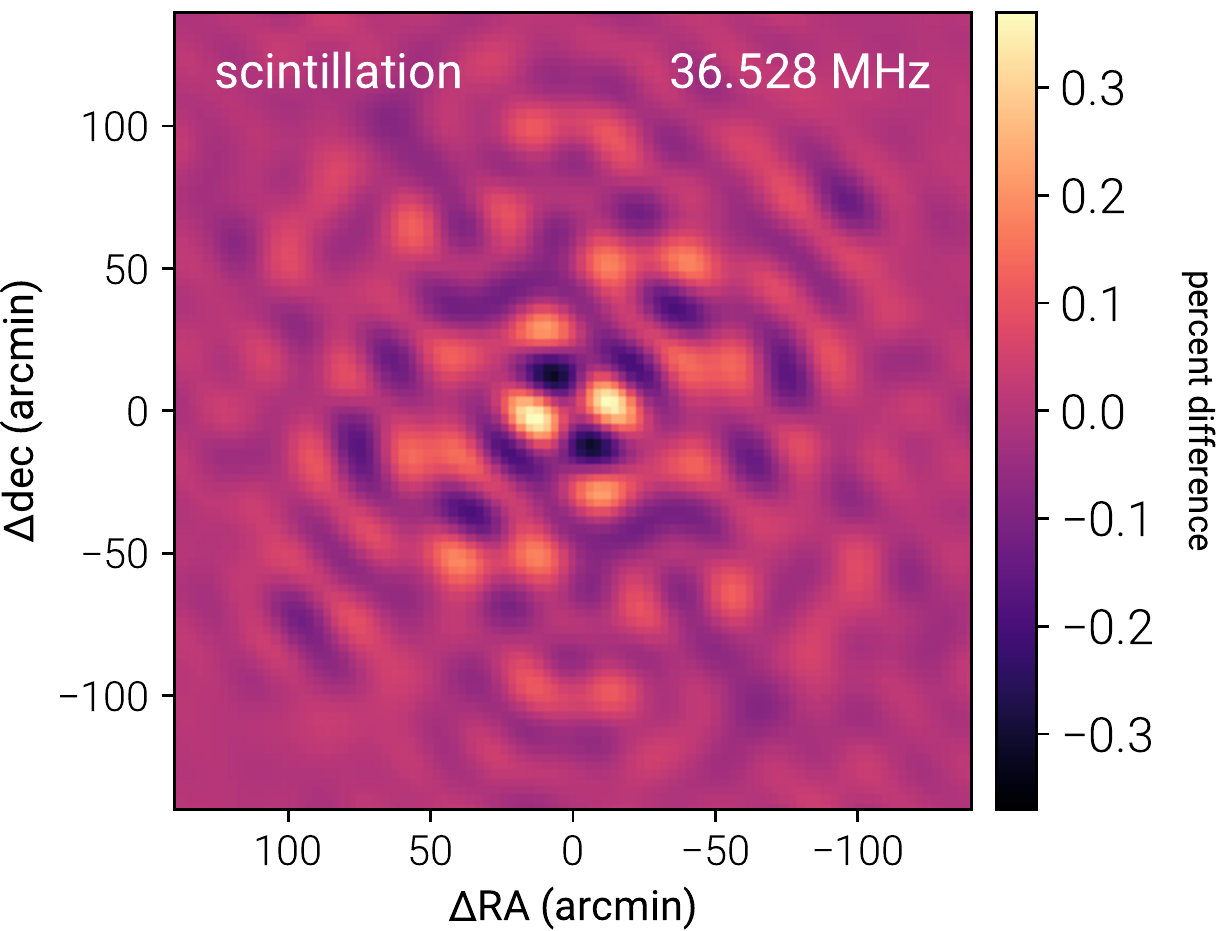} &
        \includegraphics[width=\columnwidth]{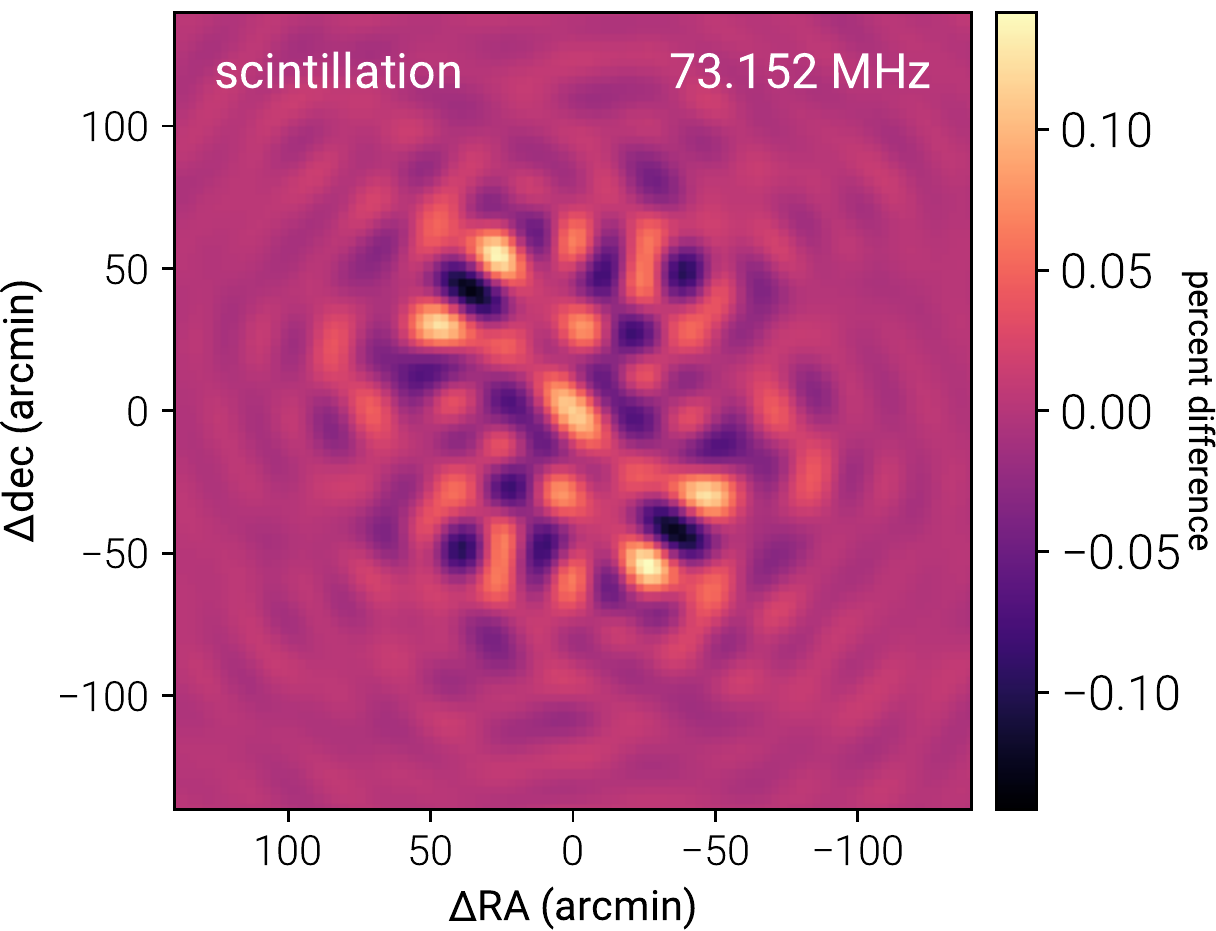} \\
        \includegraphics[width=\columnwidth]{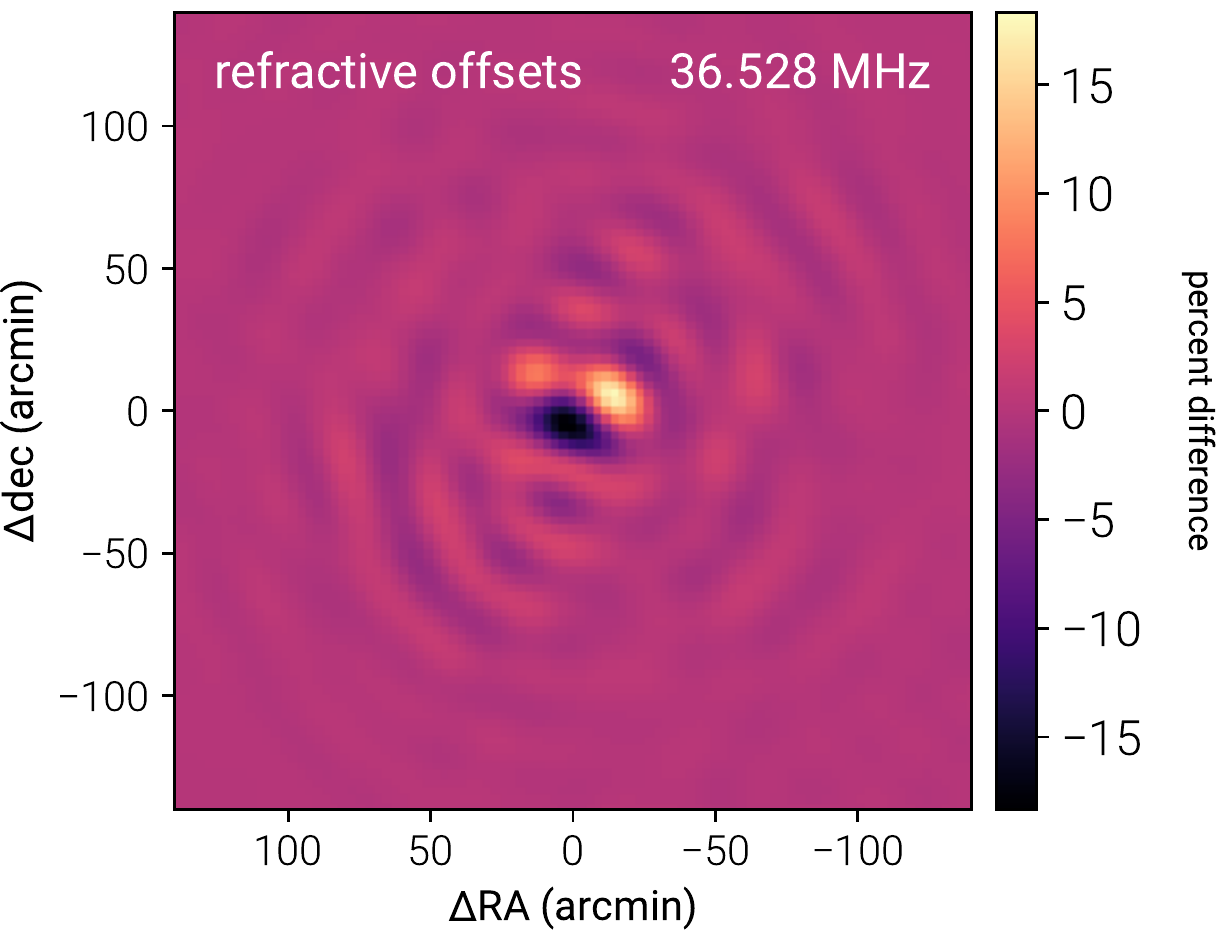} &
        \includegraphics[width=\columnwidth]{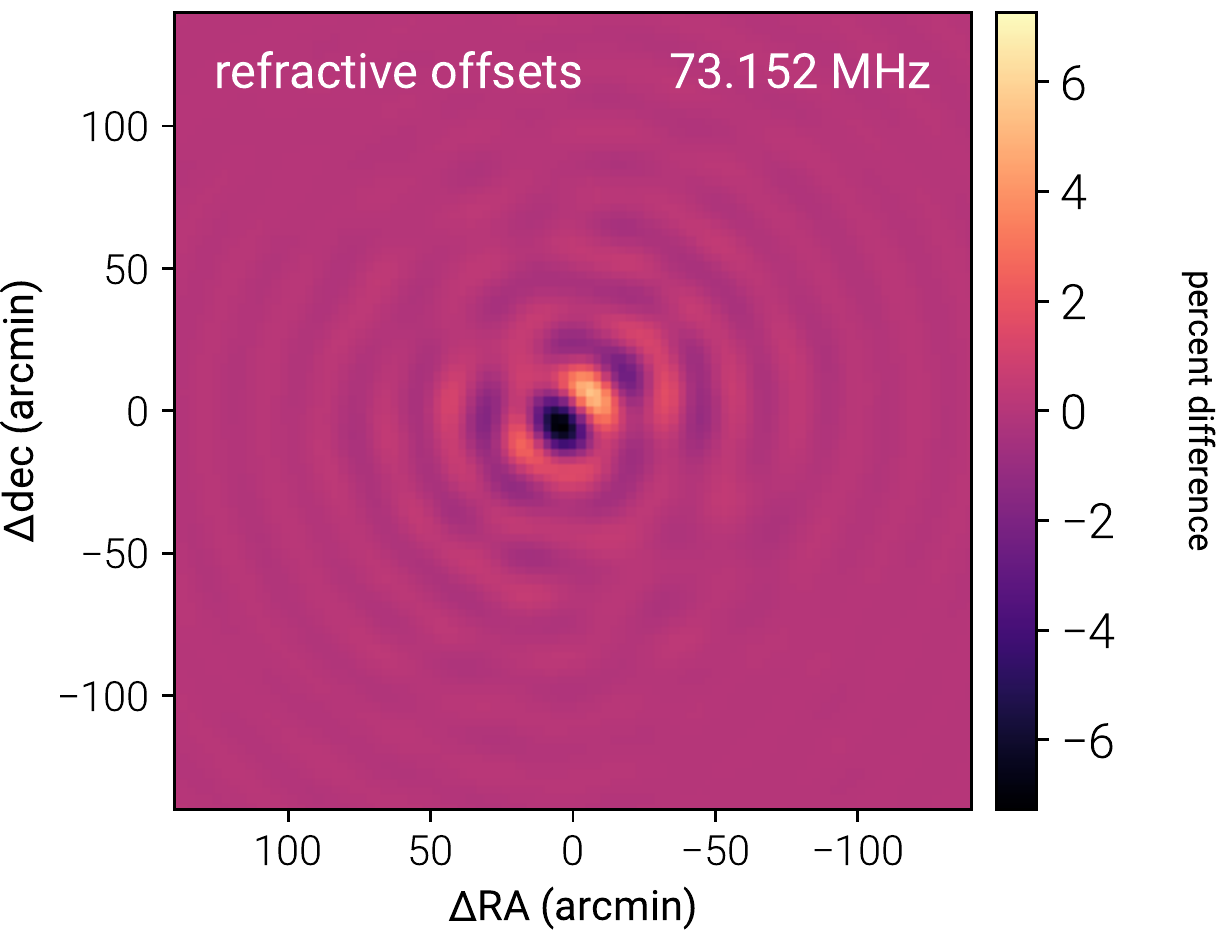} \\
    \end{tabular}
    \caption{
        Illustration of the corrupting influence of the ionosphere at 36.528~MHz (left column)
        compared with 73.152~MHz (right column). Each panel shows the \textit{simulated} PSF for a
        source at the location of Cas~A, and illustrates the percent difference (relative to the
        peak flux of the uncorrupted PSF) due to including an ionospheric effect.  In the top row,
        the simulated source scintillates using the measured light-curve for Cas~A in
        Figure~\ref{fig:scintillation}. In the bottom row, the simulated source is refracted from
        its true position using the measured refractive offsets for Cas~A in
        Figure~\ref{fig:scintillation}.
    }
    \label{fig:ionospheric-simulations}
\end{figure*}

\begin{figure*}[t]
    \centering
    \includegraphics[width=\textwidth]{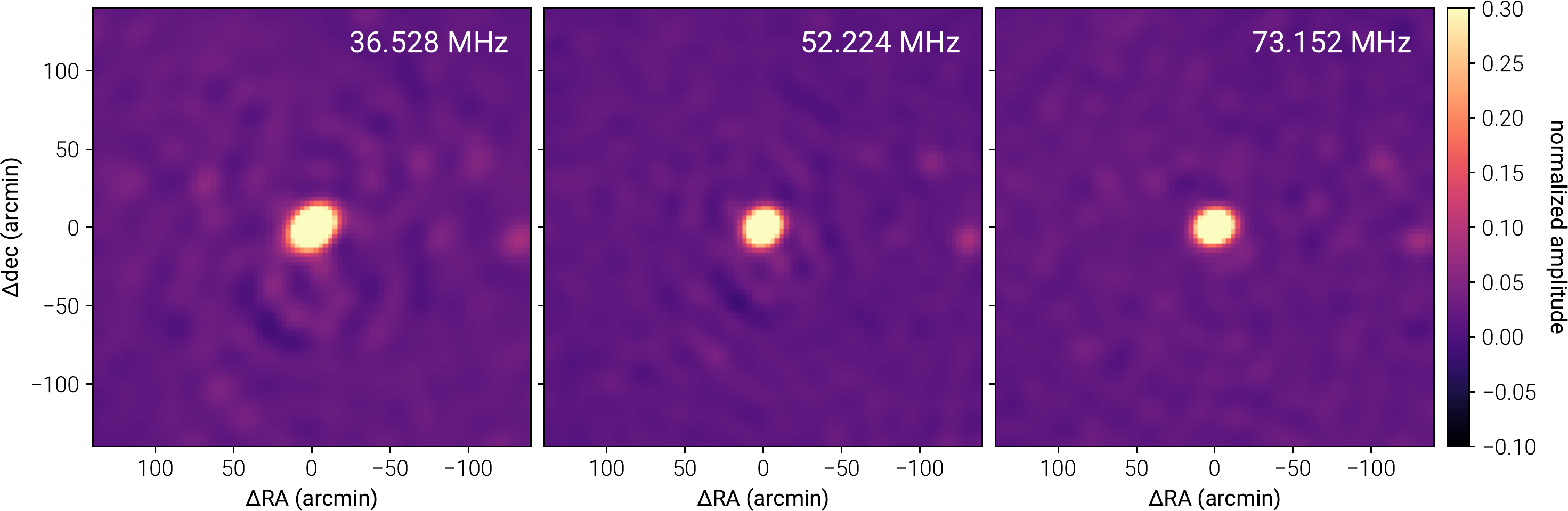}
    \caption{
        A zoom-in of 3C~134 at 36.528~MHz (left panel), 52.224~MHz (middle panel), and 73.152~MHz
        (right panel). At 36.528~MHz there are $\sim10\%$ artifacts around 3C~134 that persist after
        CLEANing due to ionospheric effects. As expected for an ionospheric origin, these artifacts
        decrease in amplitude as frequency increases. Figure~\ref{fig:ionospheric-simulations} shows
        the typically expected amplitude of these effects for ionospheric scintillation and
        refractive offsets.
    }
    \label{fig:3C134}
\end{figure*}

One of the key assumptions made by $m$-mode analysis is that the sky is static.  We assume that the
only time-dependent behavior is the rotation of the Earth, which slowly rotates the sky through the
fringe pattern of the interferometer. At low frequencies, the ionosphere violates this assumption.
In particular, ionospheric scintillation and refractive offsets will cause even static sources to
exhibit significant variability (Figure~\ref{fig:scintillation}).

The correlation observed on a given baseline for a single point source is
\begin{equation}
    V_\nu(t_{\textrm{sidereal}}) = I_\nu B_\nu(t_{\textrm{sidereal}}),
\end{equation}
where $I_\nu$ is the flux of the source at the frequency $\nu$, and $B_\nu$ is the baseline transfer
function defined by Equation~\ref{eq:baseline-transfer-function}. The transfer function is a
function of the direction to the source, which is in turn a function of the sidereal time
$t_{\textrm{sidereal}}$. If the source is varying, from intrinsic variability or due to
scintillation, than the source flux is also a function of the time coordinate $t$ such that
\begin{equation}
    V_\nu(t_{\textrm{sidereal}}) = I_\nu(t) B_\nu(t_{\textrm{sidereal}}),
\end{equation}
where $t_{\textrm{sidereal}} = (t \mod 23^h56^m)$.

In order to compute the $m$-modes we must take the Fourier transform with respect to the sidereal
time. As a consequence of the Fourier convolution theorem we find:
\begin{equation}\label{eq:no-longer-block-diagonal}
    V_{\nu, m} \sim \sum_{m^\prime} V_{m^\prime}^\textrm{static} I_{\nu, m-m^\prime}\,,
\end{equation}
where $V_{\nu, m}^{\textrm{static}}$ is the set of observed $m$-modes if the source was actually
static, and $I_{\nu, m-m^\prime}$ is the Fourier transform of the light curve $I_{\nu}(t)$.
Equation~\ref{eq:no-longer-block-diagonal} indicates that power is scattered between different
values of $m$. As a consequence, the true transfer matrix, which is exactly block diagonal in the
ideal case, is no longer truly block diagonal \citep{richard_ionosphere_thoughts}.

The maps presented in Figure~\ref{fig:channel-maps} do not account for any off-diagonal terms
arising from ionospheric fluctuations. The effect of this can be seen in
Figure~\ref{fig:ionospheric-simulations}. In this simulation, a point source is placed at the
location of Cas~A. In one case the source is allowed to scintillate in the same way Cas~A does in
Figure~\ref{fig:scintillation}, but the source is always located exactly at the location of Cas~A.
In the second case the source position is allowed to vary in the same way Cas~A does in
Figure~\ref{fig:scintillation}, but the flux of the source exactly traces the beam model. The
scintillation, although large, introduces only $<0.3\%$ errors in the vicinity of bright point
sources. Refractive offsets, however, can introduce $\sim 15\%$ errors at 36.528~MHz and $\sim 5\%$
errors at 73.152~MHz.  Because the sidelobes of the PSF are altered from that of the ideal PSF,
refractive offsets will restrict the dynamic range it is possible to obtain with the CLEAN algorithm
described in \S\ref{sec:clean}. This effect can be clearly seen in Figure~\ref{fig:3C134}, where
10\% errors within 1$^\circ$ of 3C~134 are seen at 36.528~MHz.  As expected for an ionospheric
effect these errors decrease to a few percent at 52.224~MHz, and less at 73.152~MHz. We therefore
conclude that ionospheric effects directly limit the dynamic range in the vicinity of bright point
sources.

\subsection{Beam Errors}

A model of the antenna beam is essential for widefield imaging. Because $m$-mode analysis imaging
operates on a full sidereal day of data, images are constructed after watching each point in the sky
move through a large slice of the beam (excepting the celestial poles). The beam model therefore
serves two purposes:
\begin{enumerate}
    \item setting the flux scale as a function of declination
    \item reconciling observations from two separate sidereal times
\end{enumerate}

In the first case, all sources at a given declination take the same path through the antenna primary
beam. If the antenna response is overestimated along this track, then all sources at this
declination will have underestimated fluxes. Similarly, if the antenna response is underestimated,
then all the sources will have overestimated fluxes. The errors in Figure~\ref{fig:flux-scale} do
not show a clear pattern with declination. Two sources have a clear systematic offset at all
frequencies: 3C 353 and 3C 380. 3C 353 is the second southernmost source, but Hya A -- the first
southernmost source -- does not exhibit this systematic error. Similarly, 3C 380 is at a comparable
declination to Lyn A, which appears, if anything, to have its flux systematically offset in the
other direction. The absence of a coherent pattern does not eliminate the possibility of beam errors
effecting the flux scale, but it does mean that these errors are at least comparable to the errors
inherent to the flux scale itself.

The second case is more subtle. Sources are observed at a wide range of locations in the primary
beam of the antenna. The imaging process must reconcile all of these observations together, and the
beam model provides the instructions for how to do this. In the event of an error in the beam model,
it can be expected that the beam will introduce errors into the sky maps that will limit the dynamic
range in the vicinity of bright point sources.  \citet{2015PhRvD..91h3514S} simulate the effect of
beam errors on a cosmological analysis concluding that the beam must be known to one part in $10^4$.
Our requirements are significantly less stringent because we are estimating the sky brightness
instead of estimating the amplitude of a faint cosmological signal in the presence of foreground
emission that dominates the signal by five orders of magnitude. In fact, in \S\ref{sec:ionosphere}
we found that ionospheric effects likely dominate over other sources of error that affect the PSF
shape. Therefore we conclude that the beam models generated in \S\ref{sec:beam} are sufficient to
limit the effect of beam errors on the PSF to at least less than those introduced by the ionosphere.

\subsection{Polarization Leakage}

\citet{2015PhRvD..91h3514S} describe how to generalize $m$-mode analysis to account for a polarized
sky observed with a polarized antenna beam. Heretofore this generalization has been neglected in the
discussion of $m$-mode analysis imaging.  At low frequencies, increasingly rapid Faraday rotation
leads to depolarization. Therefore polarization fractions are generally expected to decrease at low
frequencies (varying with ionospheric conditions). \citet{2016ApJ...830...38L} detected the presence
of diffuse polarized emission on degree angular scales with the MWA, also finding typical
depolarization ratios of $\sim0.3$ for pulsars at 154~MHz relative to 1.4~GHz, although there was a
large variance between pulsars. Even more depolarization is expected at frequencies $\le
73.152$~MHz, but crossed-dipole antennas with extremely large primary beams will naturally introduce
large polarization leakage terms at low elevations.  It is instructive to compute what impact this
will have on the unpolarized imaging process.

In order to understand the effect of polarization leakage, we simulated a point source with 10\%
polarization in Stokes-$Q$ at the location of Cas A.  The simulated visibilities were computed using
the measured beams for $x$- and $y$-dipoles. Because the amplitude of the two beams are not equal in
every direction on the sky, this introduces a direction-dependent leakage of Stokes-$Q$ into
Stokes-$I$. At 73.152~MHz, this leakage is $\lesssim5\%$ above $15^\circ$ elevation, but rapidly
rises to $\gtrsim50\%$ at lower elevations. \citet{2015JAI.....450004O} report similar polarization
leakage measurements with the LWA1.  Cas A is a circumpolar source and spends about seven hours
every day skirting the horizon where the polarization leakage is large, so by placing the simulated
source at the location of Cas A, we are engineering a situation where the polarization leakage from
Stokes-$Q$ into Stokes-$I$ will be large. However, the impact on the unpolarized $m$-mode analysis
maps is mild, amounting to a 0.5\% error in the flux of the source with no measurable effect on the
PSF.

\subsection{Terrestrial Interference and Pickup}\label{sec:rfi}

\begin{figure*}[t]
    \includegraphics[width=\textwidth]{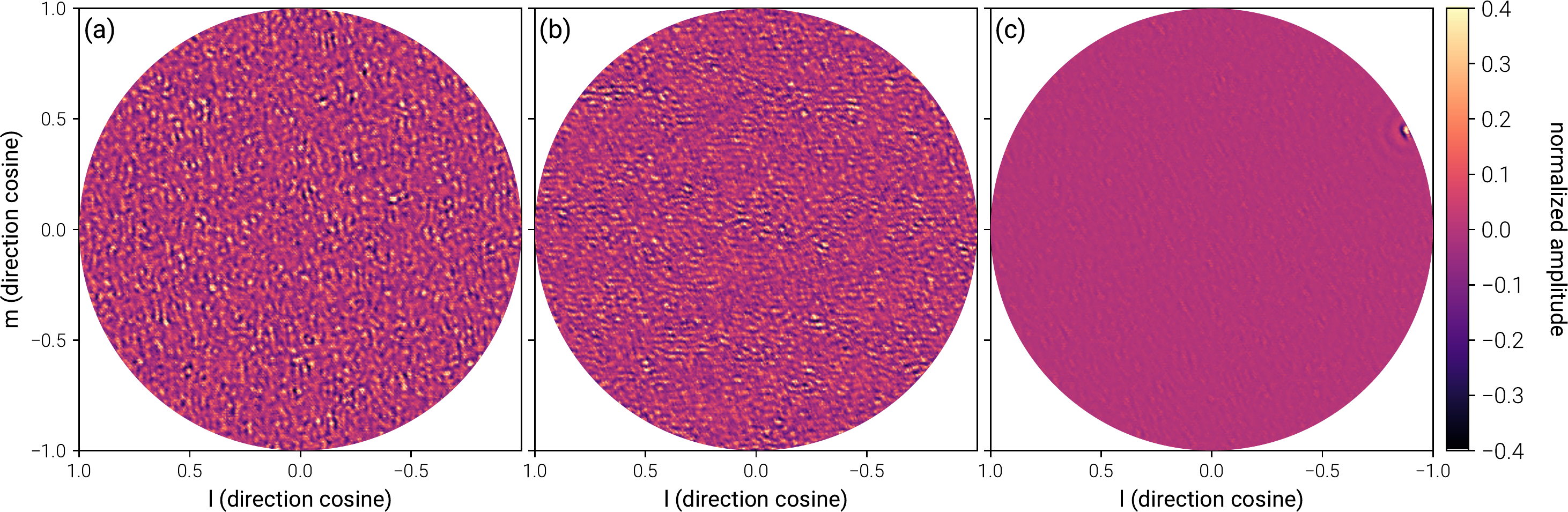}
    \caption{
        Terrestrial sources of correlated noise that are apparent after averaging the visibilities
        at 62.688 MHz over the entire 28 hour observing period (keeping the phase center at zenith
        such that astronomical sources of radio emission are smeared along tracks of constant
        declination). Each panel represents a different component that is removed from the
        visibilities. The images are generated using WSClean \citep{2014MNRAS.444..606O}, uniform
        weighting, and only baselines longer than 15 wavelengths. Panels (a) and (b) illustrate
        components that appear noise-like in image-space, but are in fact a constant-offset to the
        measured visibilities likely associated with cross-talk or common-mode pickup. Panel (c)
        illustrates a component that is clearly associated with an RFI source on the horizon to the
        west-north-west of the OVRO-LWA. This RFI source is likely an arcing power line.
        Figure~\ref{fig:rings} illustrates the characteristic ring-like artifacts introduced into
        the maps if these 3 components are not removed prior to $m$-mode analysis imaging. The
        component shown in panel (a) has about twice the amplitude ($\|\b v_\text{terrestrial}\|$)
        of panels (b) and (c), and for all three components $\|\b B^*\b v_\text{terrestrial}\|/(\|\b
        B\|\|\b v_\text{terrestrial}\|) \sim 0.035$.
    }
    \label{fig:fitrfi}
\end{figure*}

\begin{figure*}[t]
    \centering
    \includegraphics[height=0.32\textheight]{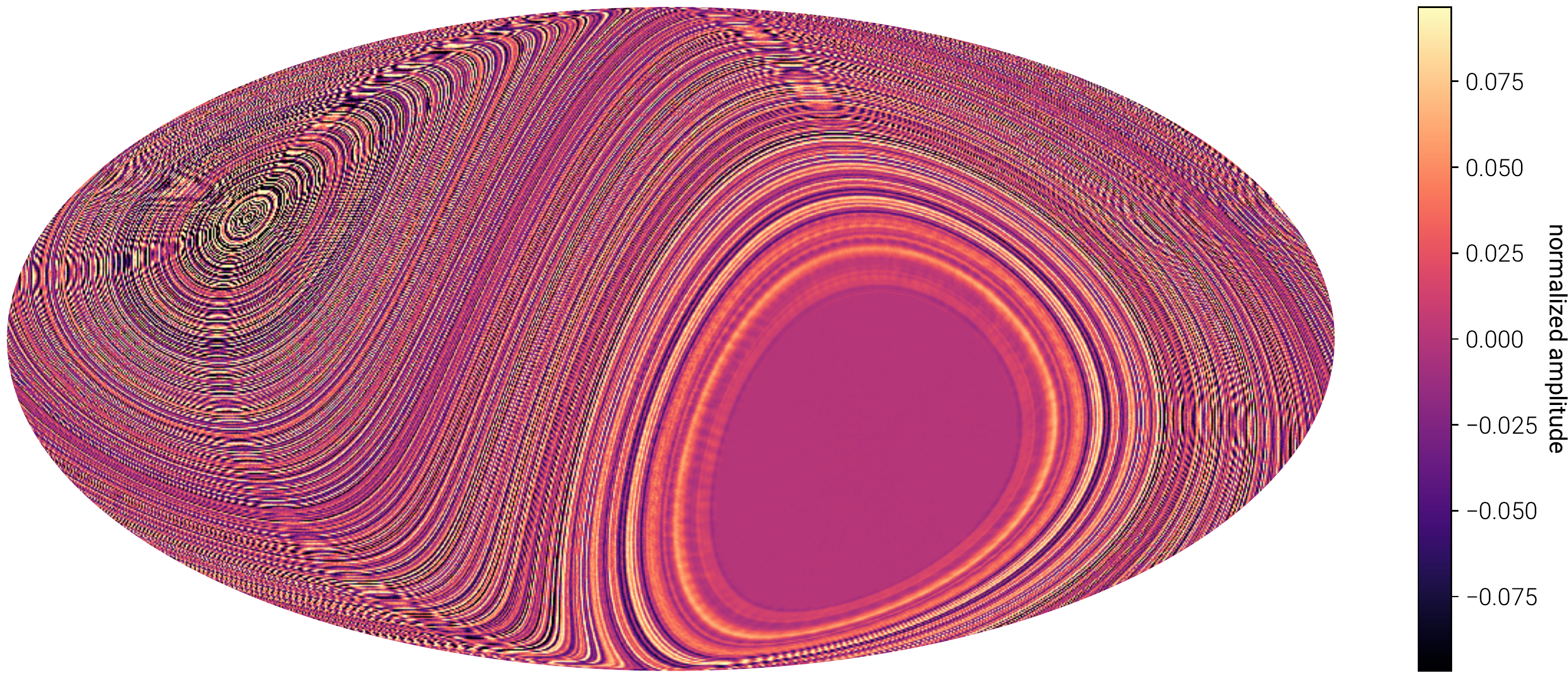}
    \caption{
        A Mollweide-projected image of the artifacts introduced to the $m$-mode analysis maps by the
        three terrestrial sources shown in Figure~\ref{fig:fitrfi}. Because these sources are not
        moving through the sky sidereally, they tend to be smeared along rings of constant
        declination. The spurs seemingly radiating from the north celestial pole are a Moir\'{e}
        pattern (ie. an artifact of the pixelization).
    }
    \label{fig:rings}
\end{figure*}

When writing Equation~\ref{eq:basic-imaging}, it is implicitly assumed that the correlated voltage
fluctuations measured between pairs of antennas are exclusively generated by astronomical sources of
radio emission. In practice, this assumption can be violated. For instance, a low-frequency
interferometer located in the vicinity of an arcing power line will see an additional contribution
from the radio-frequency interference (RFI) generated by the arcing process. Similarly, common-mode
pickup along the analog signal path of the interferometer may generate an additional spurious
contribution to the measured visibilities. While the amplitude and phase of these contaminating
signals may fluctuate with time, they do not sweep across the sky at the sidereal rate
characteristic of astronomical sources.

The Owens Valley is an important source of water and power for the city of Los Angeles.
Unfortunately, this means that high voltage power lines run along the valley $\gtrsim10$~km to the
west of the OVRO-LWA. Some of these power line poles have faulty insulators that arc and produce
pulsed, broadband RFI. Because these poles exist in the near-field of the array, we have been able
to localize some of them by using the curvature of the incoming wavefront to infer a distance.
Efforts are currently underway to work with the utility pole owners to have these insulators
replaced.

In the meantime, it is possible to suppress their contamination in the dataset. The contribution of
these RFI sources to the visibilities can be plainly seen by averaging $>24$ hours of data with the
phase center set to zenith. In this way, true sky components are smeared along tracks of constant
declination while terrestrial sources (ie. the arcing power lines or any contribution due to
common-mode pickup) are not smeared.  Obtaining a model for the RFI is complicated by the fact that
the contaminating sources are at extremely low elevations where the antenna response is essentially
unknown (and inhomogeneous due to antenna shadowing effects). It is not enough to know the physical
location of the faulty insulator generating the RFI. In addition, we must know the response of each
antenna (amplitude and phase) in the appropriate direction. This motivates the use of peeling, which
allows the antenna response to be a free parameter.  Therefore, model visibilities for the RFI can
be obtained by peeling the sources after smearing the visibilities over $>24$ hours.
Figure~\ref{fig:fitrfi} shows an illustration of some of the removed components at 62.688 MHz.

While attempting to peel RFI sources from the averaged visibilities, it was discovered that
frequently peeling would remove components from the visibilities that are not obviously associated
with any source on the horizon or elsewhere in the sky (see panels (a) and (b) in
Figure~\ref{fig:fitrfi}).  These components appear noise-like in the images, but they are actually a
constant offset to the measured visibilities and are therefore likely associated with cross-talk or
some form of common-mode pickup. If these components are not subtracted from the measured
visibilities, they contribute ring-like structures into the sky maps as seen in
Figure~\ref{fig:rings}. This figure is not a simulation, but rather a difference between maps
created before and after measuring and subtracting the components in Figure~\ref{fig:fitrfi} from
each integration.

The first step in Equation~\ref{eq:tikhonov-solution} is to compute $\b B^*\b v$. In this step we
compute the projection of the measurement $\b v$ onto the space spanned by the columns of $\b B$.
Each column of $\b B$ describes the interferometer's response to a corresponding spherical harmonic
coefficient of the sky brightness distribution. Therefore, the act of computing $\b B^*\b v$ is to
project the measured $m$-modes onto the space of $m$-modes which could be generated by astronomical
sources. The degree to which a source of terrestrial interference will contaminate a map generated
using $m$-mode analysis imaging is determined by its amplitude after projection.

For instance, a bright interfering source might contribute $\b v_\text{terrestrial}$ to the measured
$m$-modes. However, if $\b v_\text{terrestrial}$ is actually perpendicular to all of the columns of
$\b B$, there will be no contamination in the map because $\b B^*\b v_\text{terrestrial} = \b 0$.
In practice this is unlikely. In general, the contamination is proportional to the overall amplitude
of the interference ($\|\b v_\text{terrestrial}\|$) and the degree to which the interference mimics
an astronomical signal ($\|\b B^*\b v_\text{terrestrial}\|/(\|\b B\|\|\b v_\text{terrestrial}\|)$).

These terrestrial sources do not rotate with the sky and hence their contamination tends to be
restricted to modes with small $m$. In this dataset the contamination is largely restricted to $m
\lesssim 1$. Although the RFI is capable of fluctuating on short timescales, in this case the
artifacts it introduces seem to be restricted to small $m$ (presumably because the phase is not
fluctuating).  As a result if the contamination is not suppressed, it will manifest itself as rings
along stripes of constant declination. This effect is plainly visible in Figure~\ref{fig:fitrfi}.
Because of the distinctive ring-like pattern created by terrestrial sources, we additionally chose
to discard spherical harmonics with either $m=0$ or $m=1$ and $l>100$ in order to further suppress
the contamination.

\section{Conclusion}\label{sec:conclusion}

In this work we presented a new imaging technique -- Tikhonov regularized $m$-mode analysis imaging
and CLEANing -- for drift-scanning telescopes like the OVRO-LWA.  This technique exactly corrects
for widefield effects in interferometric imaging with a single synthesis imaging step.  We applied
Tikhonov regularized $m$-mode analysis imaging to a 28 hour dataset and generated eight sky maps
between 36.528~MHz and 73.152~MHz.  These sky maps are a substantial improvement in angular
resolution over existing maps at these frequencies with $\sim$15~arcmin angular resolution and
$<600$~K thermal noise. The point source flux scale is consistent with that defined by
\citet{2012MNRAS.423L..30S} to about 5\% and large angular scales are consistent with the work of
\citet{2017MNRAS.469.4537D} to within 20\%.

At frequencies above $\sim55$~MHz, the angular resolution of these maps is limited by the selection
of $l_\text{max}=1000$. Future work will increase $l_\text{max}$ to remove this restriction, as well
as include more time and bandwidth to improve the thermal noise. The usage of nighttime-only data
can help mitigate dynamic range limitations from the ionosphere and also eliminate solar sidelobe
residuals. Observations could also be extended to slightly higher and lower frequencies ($\sim27$ to
$85$~MHz) to take advantage of the full frequency range of the OVRO-LWA. The higher frequencies are
particularly interesting in order to maximize the overlap with the MWA in the southern hemisphere,
which could be used to fill-in the hole around the southern celestial pole.

These maps and future improvements are primarily intended to be used as part of a foreground
modeling and subtraction routine for 21-cm cosmology experiments. Each map will be made publicly
available on LAMBDA.

\acknowledgments
This work is dedicated to the memory of Professor Marjorie Corcoran, who was an influential mentor
to MWE.

This material is based in part upon work supported by the National Science Foundation under Grant
AST-1654815 and AST-1212226. The OVRO-LWA project was initiated through the kind donation of Deborah
Castleman and Harold Rosen.

Part of this research was carried out at the Jet Propulsion Laboratory, California Institute of
Technology, under a contract with the National Aeronautics and Space Administration, including
partial funding through the President's and Director's Fund Program.

This work has benefited from open-source technology shared by the Collaboration for Astronomy Signal
Processing and Electronics Research (CASPER).  We thank the Xilinx University Program for donations;
NVIDIA for proprietary tools, discounts, and donations; Digicom for collaboration on manufacture and
testing of DSP processors.

We thank the Smithsonian Astrophysical Observatory Submillimeter Receiver Lab for the collaboration
of its members.

Development, adaptation, and operation of the LEDA real-time digital signal processing systems at
OVRO-LWA has been supported in part by NSF grants AST/1106059, PHY/0835713, and OIA/1125087.

GBT, JD and FKS acknowledge support from the National Science Foundation under grant AST-1139974.

\bibliographystyle{aasjournal}
\bibliography{paper}

\end{document}